\documentclass[aoas, preprint]{imsart}

\RequirePackage{amsthm,amsmath,amsfonts,amssymb}
\RequirePackage[authoryear]{natbib}
\RequirePackage[colorlinks,citecolor=blue,urlcolor=blue]{hyperref}
\RequirePackage{graphicx}

\usepackage{subcaption}
\usepackage{bm}
\usepackage{bbm}
\usepackage{booktabs}
\usepackage{threeparttable}
\usepackage{adjustbox}
\usepackage{siunitx}

\startlocaldefs
\theoremstyle{plain}

\theoremstyle{definition}

\endlocaldefs

\begin{document}

\begin{frontmatter}
\title{Dynamic Count Models with Flexible Innovation Processes for Irregular Maritime Migration}
%\title{A sample article title with some additional note\thanksref{t1}}
\runtitle{Dynamic Count Models for Maritime Migration}
%\thankstext{T1}{A sample additional note to the title.}

\begin{aug}
%%%%%%%%%%%%%%%%%%%%%%%%%%%%%%%%%%%%%%%%%%%%%%%
%% Only one address is permitted per author. %%
%% Only division, organization and e-mail is %%
%% included in the address.                  %%
%% Additional information can be included in %%
\author[A]{\fnms{Gregor}~\snm{Zens}\ead[label=e1]{zens@iiasa.ac.at}}
\and
\author[B]{\fnms{Jakub}~\snm{Bijak}\ead[label=e2]{jakub.bijak@demography.ox.ac.uk}}\\
\vspace{1em}
%\textcolor{red}{[Preliminary, please do not distribute or publish without permission.]}

%%%%%%%%%%%%%%%%%%%%%%%%%%%%%%%%%%%%%%%%%%%%%%
%% Addresses                                %%
%%%%%%%%%%%%%%%%%%%%%%%%%%%%%%%%%%%%%%%%%%%%%%
\address[A]{Population and Just Societies Program, International Institute for Applied Systems Analysis (IIASA)\printead[presep={,\ }]{e1}}

\address[B]{Leverhulme Centre for Demographic Science, Nuffield Department of Population Health \\and Reuben College, University of Oxford\printead[presep={\ }]{e2}}
\end{aug}

\begin{abstract}

Motivated by the challenge of analyzing the dynamics of weekly sea border crossings in the Mediterranean (2015--2025) and the English Channel (2018--2025), we develop a Bayesian dynamic framework for modeling heteroskedastic count time series. Building on theoretical considerations and empirical stylized facts, our approach utilizes a Poisson random walk model that allows for heavy-tailed innovations or stochastic volatility dynamics, while incorporating an explicit mechanism to separate structural from sampling zeros. Posterior inference is carried out via a straightforward Markov chain Monte Carlo algorithm. Applying this methodology to Mediterranean and English Channel data, we compare alternative model specifications through a comprehensive out-of-sample forecasting exercise. Using log predictive scores and empirical coverage at predictive quantiles to evaluate each model, we find strong evidence for stochastic volatility in migration innovations. These models deliver the strongest out-of-sample forecasts with empirical coverage close to nominal levels up to the 99th percentile. Our framework can be used to develop risk indicators with direct policy implications for improving governance and preparedness for migration surges. More broadly, the methodology extends to other zero-inflated non-stationary count time series applications, including epidemiological surveillance and public safety incident monitoring.

\end{abstract}


\end{frontmatter}
%%%%%%%%%%%%%%%%%%%%%%%%%%%%%%%%%%%%%%%%%%%%%%
%% Please use \tableofcontents for articles %%
%% with 50 pages and more                   %%
%%%%%%%%%%%%%%%%%%%%%%%%%%%%%%%%%%%%%%%%%%%%%%
%\tableofcontents

\section{Introduction}

Irregular maritime migration and sea border crossings are major issues in contemporary political discourse and come with severe humanitarian consequences due to the high mortality risks associated with sea migration routes. At the same time, such "small boat" passages constitute one of the few irregular migration contexts where arrivals can be tracked with relative precision, based on systematic search-and-rescue logs, and immigration or asylum registration systems in arrival countries. The resulting availability of high-frequency data should, in principle, enable a rigorous statistical investigation. 

However, the literature on modeling the dynamics of irregular maritime migration and border crossings remains surprisingly thin. This gap may stem from two intertwined obstacles. First, data on irregular arrivals by sea present unique challenges, as high-frequency counts of migrant arrivals are noisy, non-stationary, volatile, and often zero-inflated -- a combination that overwhelms many off-the-shelf modeling techniques. Second, these characteristics make point forecasts notoriously unreliable, potentially creating the impression that any attempt to effectively model irregular maritime border crossings is futile. This leaves a significant research gap both in terms of forecasting capabilities that could enhance preparedness and humanitarian response, and in terms of developing structural insights into the underlying drivers and dynamics of irregular maritime migration.

In this article, we demonstrate that both meaningful structural insights and well-calibrated distributional forecasts of sea border crossing activity can be obtained using a simple context-specific probabilistic model. Based on theoretical considerations and stylized empirical facts of irregular maritime border crossing data, we develop a Bayesian dynamic modeling framework for count data time series. This framework is designed for discrete time series jointly characterized by (i) highly persistent trending patterns, (ii) volatility clustering or sudden shifts in the log-scale intensity, and (iii) zero-inflation. We develop a simple Markov chain Monte Carlo (MCMC) algorithm for posterior simulation. The proposed methods and algorithms are flexible, easily extendable, and applicable to any non-negative integer time series. These features suggest their broader utility across various fields beyond modeling irregular sea border crossings.

We apply the model to two case study datasets on weekly maritime border crossings in the Mediterranean Sea and in the English Channel. We benchmark various versions of the model to gain insight into irregular migration dynamics, focusing particularly on the problem of producing and evaluating distributional forecasts. Specific emphasis is placed on empirical coverage in the tails of the predictive distribution of maritime border crossings. This focus is especially relevant, as accurate tail probability forecasts directly enhance preparedness for extreme events. For instance, well-calibrated predictive distribution tails may enable more effective contingency planning in the context of irregular migration, which is often characterized by rare events of high magnitude \cite[see e.g.][]{bijak2024}. Such approaches are already entering the mainstream debate and policy in other areas, such as preparedness for extreme climate events \cite[e.g.][]{ipcc2012}, or civil contingencies more generally \citep{riskreg25}. The emerging frameworks for responding to high-impact migration events (e.g. the EU \textit{Migration Preparedness and Crisis Blueprint}, \citeauthor{ec2020}, \citeyear{ec2020}) open the possibility of extending such approaches to migration in a formal and rigorous way.

Our results indicate that the dynamics of irregular sea border crossings are effectively captured by a Poisson state-space model with log intensity expressed as a random walk with stochastic volatility. In both our case studies, this model delivers the strongest one-week-ahead probabilistic forecasting performance for irregular sea border crossing activity among the specifications considered, with empirical coverage close to nominal levels at upper-tail quantiles such as $q_{95}$ and $q_{99}$. We connect these findings with the underlying theoretical considerations and discuss implications for policy and future empirical research under the assumption that sea border crossings are indeed generated by the proposed model.

This article therefore offers two primary contributions. First, from a statistical perspective, we introduce a flexible modeling framework for analyzing zero-inflated and potentially non-stationary count time series, accompanied by a straightforward computational algorithm for posterior simulation. This advances the literature on Bayesian approaches to dynamic modeling of complex integer-valued time series. Second, in terms of substantive application, we conduct the first comprehensive predictive evaluation to determine effective approaches for probabilistic modeling of irregular maritime border crossing data through two high-profile and policy-relevant case studies. This yields a new benchmark model for examining such processes and generates evidence-based policy insights regarding the underlying structural dynamics.

The remainder of the article is structured as follows. Section~\ref{sec:data} presents the motivating case study datasets, empirical stylized facts, theoretical considerations, and related literature. Section~\ref{sec:model} presents our model, prior specification, and the computational approach. Section~\ref{sec:application} applies the methodology to the case study datasets. Section~\ref{sec:conclusion} discusses the policy implications of the results and concludes with directions for future research. %A simulation study and additional results are presented in the supplementary material (Zens and Bijak, 2026).

\section{Data, Stylized Facts, and Theoretical Considerations}
\label{sec:data}

We analyze weekly time series data on irregular sea border crossings across two distinct geographical regions. The first case study examines irregular arrivals in Italy across the Mediterranean Sea, encompassing $T=494$ weeks from 2015W40 to 2025W11. These data are sourced from the \textit{Operational Data Portal} of the United Nations High Commissioner for Refugees (UNHCR).\footnote{Source: \url{https://data.unhcr.org/en/situations/europe-sea-arrivals/location/24521} (as of 7 August 2025).} On average, we observe 1,607 arrivals per week, with a weekly median of 851 arrivals. The series ranges from zero arrivals up to a maximum of 15,694 arrivals in the week of August 29, 2016.

For our second case study, we investigate irregular arrivals by sea to the United Kingdom across the English Channel, covering $T=376$ weeks from 2018W01 to 2025W11. This dataset is provided by the Border Force branch of the UK Home Office\footnote{Source: \url{https://www.gov.uk/government/publications/migrants-detected-crossing-the-english-channel-in-small-boats/migrants-detected-crossing-the-english-channel-in-small-boats-last-7-days} (as of 7 August 2025).}. On average, we observe 413 arrivals per week, with a weekly median of 152 arrivals. The series ranges from zero arrivals up to a maximum of 3,564 arrivals in the week of August 22, 2022.

\subsection{Empirical Stylized Facts}

\begin{figure}
     \centering
     \begin{subfigure}[b]{0.49\linewidth}
         \centering
         \includegraphics[width=\linewidth]{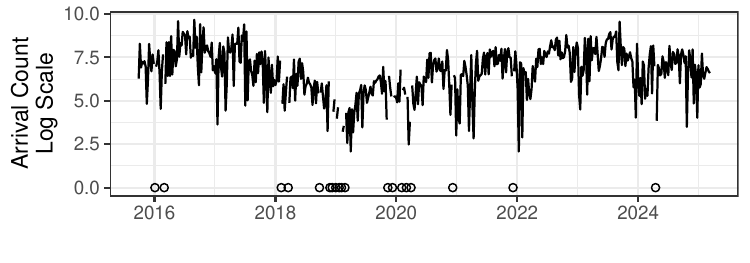}
     \end{subfigure}
          \begin{subfigure}[b]{0.49\linewidth}
         \centering
         \includegraphics[width=\linewidth]{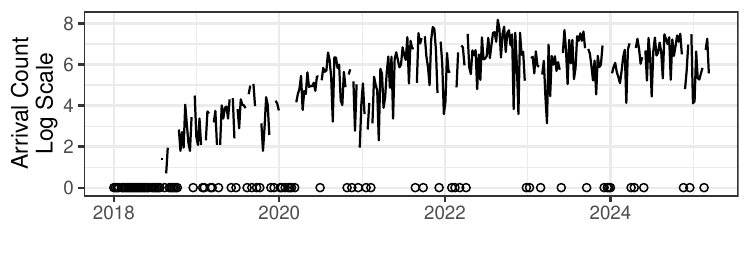}
     \end{subfigure}\\
     \begin{subfigure}[b]{0.49\linewidth}
         \centering
         \includegraphics[width=\linewidth]{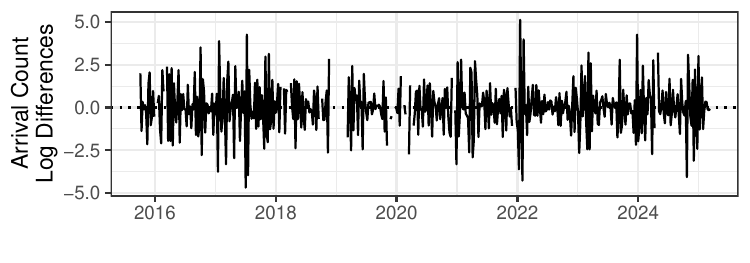}
     \end{subfigure}
     \begin{subfigure}[b]{0.49\linewidth}
         \centering
         \includegraphics[width=\linewidth]{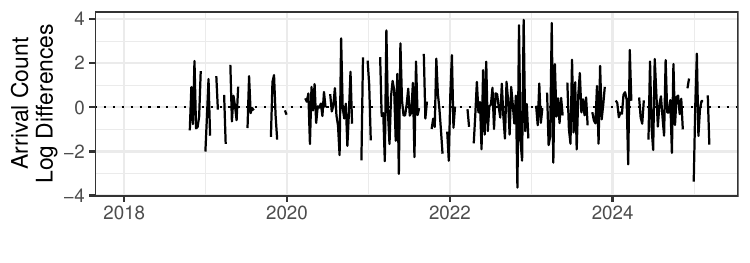}
     \end{subfigure}
     
        \caption{Irregular maritime border crossing case study datasets. Left: Mediterranean Sea border crossings to Italy by sample week from 2015W40 to 2025W11. Right: English Channel crossings by sample week from 2018W01 to 2025W11. Top panels display log-transformed counts, with zero counts highlighted separately as dots on the zero line. Bottom panels show log differences of counts for observations where both $y_t>0$ and $y_{t-1}>0$.}
        \label{fig:data_case_study}
\end{figure}

Fig.~\ref{fig:data_case_study} presents the data after a logarithmic transformation (top panels) and after taking first differences of the log-transformed counts (bottom panels). The untransformed count time series are shown in Fig.~\ref{fig:data_levels} in the supplementary material. Fig.~\ref{fig:data_case_study} highlights the three main challenges inherent in capturing boat migration time series dynamics.

First, as is common in many migration processes \citep{bijak2010}, the count-valued data exhibit high persistence with respect to mean intensity. While crossing conditions typically lead to somewhat higher activity in summer months than in winter, there is no clear deterministic trend over the full sample period. To capture this behavior, we will specify a benchmark count time series model that allows for random walk behavior of the log intensity parameter. The simple random walk assumption is empirically further motivated by the log differences in the bottom panels in Fig.~\ref{fig:data_case_study} being almost exclusively centered around the zero line, suggesting the absence of a deterministic drift.

Second, as becomes evident from the time series of log differences, the innovation process on the log scale is heteroskedastic. A priori, it is unclear whether this heteroskedasticity is the result of an overall heavy-tailed (but constant in time) process, or whether volatility itself follows a dynamic process that produces persistent time-varying volatilities that may be clustered in time. To investigate this, we will compare various volatility models empirically by benchmarking them against each other in a large-scale pseudo-out-of-sample forecasting exercise.

Third, observed zero counts may reflect one of two distinct phenomena. On the one hand, zero arrivals may be the result of weeks where external circumstances -- such as coast guard operations, adverse weather conditions, or low smuggler availability -- temporarily prevent crossings despite relatively high migration intensity in surrounding weeks (compare the zero counts after 2022 in the UK time series). In other cases, zero counts may stem from genuinely low migration intensity during those periods (such as in the early period of the UK data). There is no clear \textit{a priori} way to classify a zero as an unexpected ``outlier'' which can be treated as missing when modeling migration intensity versus treating a zero as an important data point that informs the model about low migration activity in certain time periods. This motivates an approach that treats the probability of crossing and the intensity of the underlying migration process as two separate processes that jointly generate the observed arrival counts. This conceptualization naturally leads to a parameter-driven time series model for zero-inflated counts.

\subsection{Theoretical Considerations}

The empirical patterns observed in Fig.~\ref{fig:data_case_study} -- namely high persistence, heteroskedasticity, and zero-inflation -- align with what one would expect from irregular maritime migration processes, which emerge from the complex interaction of individual decisions, weather conditions, smuggling operations, border enforcement measures (including pushbacks and deportations), political and diplomatic developments, and numerous other factors \citep[see e.g.][]{rodriguez2023search,frontex,migobs}. Many of these migration determinants can be conceptualized as evolving approximately as first-order random walks with time-varying innovation variances. Such dynamics represent plausible assumptions for classical migration drivers, including conflict pressures, environmental disasters, macroeconomic developments, smuggler capacity, and border enforcement efforts.\footnote{Note that in many cases, the direct effects of these drivers may be difficult to model and quantify explicitly. In such instances, random walk models have been proposed as atheoretical alternatives for analyzing migration when precise knowledge of drivers is unavailable \citep{bijak2010}.} Adopting a simplified perspective, one may reasonably assume that the dynamic properties of these migration drivers are inherited by the migration process itself. This process is then punctuated by occasional events -- such as coast guard operations or exceptionally bad weather conditions -- that explain the zero-inflation by temporarily halting all movement. While these considerations connect our three stylized facts, and ultimately our statistical model, to basic theoretical intuition, a fully formalized and consistent micro-theoretical treatment of irregular sea migration remains largely absent from the literature. The development of such a framework is left for future research.

Note that many of the underlying migration determinants and drivers are shared between irregular maritime migration and irregular land border crossings. Consequently, a random walk with stochastic volatility dynamics could be expected to provide a reasonable baseline for modeling the latter as well. A formal investigation of this matter is left for future research. The primary reasons we study maritime migration as a distinct subject in the current manuscript are: (i) the availability of relatively high-frequency weekly data, which is seldom the case for land border crossings -- at least not in publicly available data sources; (ii) the zero-inflation mechanism, whereby physical barriers such as the sea render crossings impossible during certain periods, a feature that is presumably less relevant for land border crossings; and (iii) the relatively distinct public perception and impact of maritime versus land migration dynamics \citep{bhatiya2025small}.

\subsection{Existing Statistical Approaches}

Conceptual statistical work on high-frequency irregular sea migration data (and high-frequency migration patterns more generally) remains relatively scarce. The majority of quantitative literature in this domain focuses on causal inference studies that evaluate the impact of factors such as search-and-rescue operations, coast guard activity, or smuggler network behavior on migration patterns (\citealp{camarena2020political}; \citealp{hoffmann2024strategic}; \citealp{friebel2024international}; \citealp{deiana2024migrants}; \citealp{zambiasi2025externalizing}). These studies typically employ relatively simplistic regression frameworks for count data or Gaussian models for log-transformed counts, based on a few observable covariates or gravity-type models. Predictive evaluations are infrequent and, if provided, focus exclusively on point prediction accuracy.

In parallel with recent efforts to apply machine learning techniques for predicting asylum-related migration activity (\citealp{carammia2022forecasting}), few studies have attempted to forecast boat migration across various frequencies and case studies -- often focusing on Mediterranean border crossings -- using an array of machine learning and standard time series analytical tools, sometimes combined in ensembles (\citealp{georgiou2016identification}; \citealp{bosco2024machine}; \citealp{bosco2025supporting}). While these studies demonstrate some degree of point predictability in sea border crossing data, the fitted models are largely black-box approaches that provide limited evidence-based structural insights into underlying patterns. In addition, distributional forecasts are typically not considered. Bridging the gap between causal inference-oriented and prediction-oriented work, \citet{rodriguez2023search} use Bayesian time series models to generate counterfactuals for Mediterranean small boat migration activity, investigating the role of search-and-rescue operations. %Their work also employed spike-and-slab priors for variable selection, demonstrating the broader utility of Bayesian methods for migration modeling.

In substantive terms, we identify three key limitations in existing statistical approaches to irregular maritime migration. First, irregular migration irregular migration is typically modeled using frameworks similar to those used for regular migration flows (e.g., labor mobility), despite fundamentally different underlying dynamics. Even for regular migration dynamics, these conventional models can demonstrate poor predictive performance (\citealp{beyer2022gravity}). Second, model evaluation -- if any -- relies almost exclusively on point forecast metrics, neglecting probabilistic accuracy measures related to the calibration of predictive distributions. As we discuss in more detail later, such calibration is essential for uncertainty quantification and risk assessment. Third, maritime migration patterns are almost exclusively analyzed in the context of the Mediterranean Sea,\footnote{An exception is \citet{wood2025factors}, which appeared after earlier versions of this manuscript were circulated as a working paper.} limiting our understanding of whether observed dynamics are region-specific or reflect more universal patterns in irregular migration dynamics. In general, little is known about the dynamics and statistical properties of irregular maritime migration.

Our work addresses these gaps and hence makes several contributions to the literature on migration modeling. First, we construct a parsimonious probabilistic model tailored specifically to high-frequency irregular maritime migration time series, grounded in both theoretical reasoning and observed empirical patterns. Despite requiring no covariates, this framework generates accurate distributional forecasts and yields insights relevant to policy design. Second, we pioneer the evaluation of distributional forecast accuracy in this domain, with special emphasis on tail behavior, which is crucial for evaluating uncertainty and providing accurate risk assessments. Third, we provide the first predictive analysis of English Channel crossings alongside Mediterranean routes, broadening the empirical evidence base. More broadly, the dynamic modeling framework we develop extends naturally to applications beyond migration, including epidemiological analyses, accident modeling or reliability contexts.
\newpage
\section{Dynamic Zero-Inflated Heteroskedastic Count Models}
\label{sec:model}

\subsection{Model specification}

Let \(y_t\in\mathbb N_0\) denote a univariate count observed at equally-spaced time points \(t=1,\dots,T\). Each count can be a \emph{structural} zero or arise from a \emph{sampling} mechanism. Introduce the latent indicator
\(s_t\in\{0,1\}\) with interpretation \(s_t=0\)  (structural zero) and \(s_t=1\) (sampling path active). Conditionally on \(s_t\) and a latent log-intensity \(z_t\in\mathbb R\) the data-generating process is  

\begin{equation}
y_t\mid s_t,z_t \;\sim\;
\begin{cases}
\delta_{0}, & s_t=0,\\[6pt]
\mathcal{P}\bigl(\lambda_t\bigr), & s_t=1,
\end{cases}
\qquad\lambda_t=\exp(z_t), \; t=1,\ldots,T,
\end{equation}

where \(\delta_{0}\) denotes a point mass at zero.
The binary indicators are taken to be i.i.d.
\(s_t\sim\operatorname{Bernoulli}(\pi)\), \(\pi\in(0,1)\). More elaborate
dynamic or covariate-driven choices are discussed in Sec.~\ref{sec:conclusion}, but are not pursued here. Marginalizing over the indicator yields the zero-inflated Poisson
mixture
\begin{equation}
\label{eq:mix_rep}
p(y_t\mid z_t, \pi)=
\begin{cases}
(1-\pi)+\pi\exp(-\lambda_t), & y_t=0,\\[6pt]
\displaystyle \pi\frac{\lambda_t^{y_t}}{y_t!}\exp(-\lambda_t), & y_t\ge 1,
\end{cases}
\qquad \lambda_t=\exp(z_t).
\end{equation}

The latent state evolves as a random walk
\begin{equation}
    z_t = z_{t-1} + \varepsilon_t,\qquad 
\varepsilon_t \sim f_\varepsilon(\,\cdot\,;\boldsymbol\vartheta),
\quad t=1,\ldots,T,
\end{equation}
with $p(z_0) \propto 1$ as prior on the initial value and where the density \(f_\varepsilon\) is parametrized using hyperparameters \(\boldsymbol\vartheta\).\footnote{Note that the underlying factors driving sea migration are extremely challenging to model explicitly. Consider for instance only conflict as a single potential driver of irregular migration. Data on conflict is available in a timely, geocoded format, which is already a major benefit compared to measurements of other potential drivers, enabling detailed modeling of conflict-related migration patterns \citep[see e.g.][]{zensthalheimer}. At the same time, accounting for the activity of the smuggling gangs, time lag due to travel from the origin region to the point of embarkation, and the waiting time there is essentially a black box. As a result, a latent variable model is highly useful in this context.} For $f_{\varepsilon}$, we consider four distinct specifications. The homoskedastic Gaussian case $f_{\varepsilon} = \mathcal{N}(0, \sigma^2)$ is used as simple benchmark. For constant heavy-tailed densities, we consider a scaled t-density with $\nu$ degrees of freedom $f_{\varepsilon} = t_{\nu}(0, \sigma^2)$ and a scale mixture of two Gaussian densities $f_{\varepsilon} = \sum_{h=1}^{2} \eta_h~\mathcal{N}(0, \sigma^2_h\sigma^2)$ with $\eta_1 + \eta_2 = 1$. Finally, following \citet{bijak2010}, to model persistent time-varying volatility patterns, we consider a stochastic volatility model (\citealp{kastner2016dealing}) $f_{\varepsilon} = \mathcal{N}(0, \exp(h_t))$ where $h_t = \mu + \phi(h_{t-1}-\mu) + \xi_t$ and $\xi_t \sim \mathcal{N}(0, \sigma_\xi^2)$. 

The considered model is essentially a heteroskedastic non-Gaussian\footnote{For large counts, a zero-inflated log-normal model may offer a computationally more efficient alternative to a Poisson specification. We do not pursue this approach for two reasons. First, counts are not necessarily large in several applications, including the English Channel data considered here. Second, from a modeling perspective, migration data inherently consists of counts and is therefore most naturally treated as such.} state-space model.  Poisson state-space formulations and their extensions are well-known in the Bayesian literature (\citealp{west1985dynamic}; \citealp{fruhwirth2006auxiliary}; \citealp{berry2020bayesian}), and state-space models have been used in migration-related applications in a handful of studies \citep[e.g.][]{rodriguez2023search,zensthalheimer}. Poisson state-space models with a random-walk log intensity augmented with locally adaptive shrinkage or dynamic horseshoe priors have been applied to trend filtering in count series (\citealp{kowal2019dynamic}; \citealp{schafer2025locally}); differenced zero-inflated negative binomial specifications have been used to capture time-varying dispersion in high-frequency financial counts (\citealp{barra2018bayesian}); and spatially varying-dispersion frameworks have been considered in epidemiological settings (e.g.\ \citealp{mutiso2022bayesian}).  In general, coupling count-data likelihoods with state-space methods enables the construction of highly intricate, domain-specific models (e.g.\ \citealp{kim2018latent}).  Our proposed framework builds directly on these developments, offering a simple, flexible, and easily extendable approach to zero-inflated, heteroskedastic count time series with flexible innovation densities, where inference can be carried out via a straightforward MCMC algorithm. For comprehensive overviews of count time series methodology, see \citet{jung2006time} or \citet{davis2016handbook}; for a Bayesian treatment of static zero-inflated count models with overdispersion, see \citet{neelon2019bayesian}; and for recent work on forecasting zero-inflated count time series, see for example \citet{aktekin2026bayesian}.

\subsection{Prior Specification} 

Bayesian inference requires elicitation of suitable prior densities. In this article, we aim for weakly informative prior specifications to avoid our conclusions relying heavily on specific hyperparameter choices or prior beliefs. We use a uniform prior $\pi \sim \mathcal{B}(1,1)$ for the probability of observing a sampling count. The prior on the initial value of the latent log intensity is chosen to be flat $p(z_0) \propto 1$. For scale parameters, weakly informative inverse gamma priors of the form $\sigma^2 \sim \mathcal{IG}(2.5,0.5)$ are used across specifications. In the Student-t model, we will estimate the degrees of freedom parameter from the data jointly alongside the other parameters. For this, we pick a shifted exponential prior with rate $1/6$ and shifted to values larger than $3$ for $\nu$ to ensure at least the first three moments of the innovation density exist. For the mixture model weights, we assume $(\eta_1, \eta_2) \sim \text{Dirichlet}(1,1)$, allowing for conditionally conjugate posterior updates. Finally, for the stochastic volatility model, we rely on the default choices in the \texttt{stochvol} \texttt{R} package (\citealp{kastner2016dealing}) and use $\mu \sim \mathcal{N}(0, 10000)$, $(\phi+1)/2 \sim \mathcal{B}(5, 1.5)$ as well as $\sigma_\xi^2 \sim \mathcal{G}(0.5, 0.5)$.

\subsection{Posterior Simulation}
\label{sec:posteriorsim}

Let  
\[
\bm{z} = (z_{1},\dots,z_{T})^{\!\top}, 
\quad 
\bm{s} = (s_{1},\dots,s_{T})^{\!\top}
\]  
collect the latent states and inclusion indicators corresponding to the data \(y_{1},\dots,y_{T}\).  We use \(\bm{\vartheta}\) to denote all hyperparameters associated with a particular choice of \(f_{\varepsilon}\).  To sample from the joint posterior  
\[
p\bigl(\bm{z},\bm{s},\bm{\vartheta},z_{0},z_{T+1},\pi \mid \bm{y}\bigr),
\]  
we construct a simple Gibbs--Metropolis sampler that cycles through each full conditional in turn.  After choosing starting values and discarding an initial burn-in, repeated iteration yields valid draws from the target posterior. Note that we explicitly include both the initial value $z_0$ and the one-step-ahead forecast $z_{T+1}$ as unknown parameters to simplify the exposition below. 

\textbf{i. Updating \(\bigl(z_{0},\bm{z},z_{T+1}\bigr)\).}  
Consider the random-walk prior \(z_{t}=z_{t-1}+\varepsilon_{t}\) with an innovation density \(f_{\varepsilon}\) that admits a conditional-Gaussian representation.  
Let
\[
\bm z^\star = (z_0,\bm z^\top,z_{T+1})^\top
            = (z_0,z_1,\dots,z_T,z_{T+1})^\top .
\]
Under the random-walk prior with flat initial prior \(p(z_0)\propto 1\), the augmented state vector
\(\bm z^\star\) has an intrinsic Gaussian prior (\citealp{chan2009efficient}) with kernel
\[
p(\bm z^\star\mid \bm\vartheta)\;\propto\;
\exp\!\left(-\frac12\,\bm z^{\star\top}\bm P(\bm\vartheta)\bm z^\star\right),
\]
where the singular precision matrix is
\[
\bm{P}(\vartheta)=\bm{D}^{\!\top}\bm{K}(\vartheta)\bm{D},\qquad 
\bm{D}=\begin{pmatrix}
-1 &  1 &         &        &  \\
   & -1 &  1      &        &  \\
   &    & \ddots & \ddots &  \\
   &    &        & -1     & 1
\end{pmatrix}_{(T+1)\times (T+2)},
\]
and
\[
\bm{K}(\bm\vartheta)=\operatorname{diag}\!\bigl(k_{1},\dots,k_{T+1}\bigr)
\]
collects the increment precisions. The exact form of \(\bm K(\bm\vartheta)\) depends on the chosen specification for \(f_\varepsilon\) and will be detailed below. Combining this prior kernel with the Poisson likelihood contributions for those observations that are on the sampling path (\(s_t=1\)) yields
\[
p\!\bigl(z_{0},\bm{z},z_{T+1}\mid\bm{y},\bm{s},\bm\vartheta\bigr)\;\propto\;
\exp\!\left(-\frac12\,\bm z^{\star\top}\bm P(\bm\vartheta)\bm z^\star\right)
\prod_{t:\,s_t=1}\mathcal{P}\!\bigl(y_t;\exp(z_t)\bigr).
\]

Direct sampling from this joint distribution is difficult, so we proceed by single-site updates
\(p(z_t\mid \bm z_{-t},s_t,y_t,\vartheta)\) for \(t=0,\dots,T+1\), where \(\bm z_{-t}\) collects all latent log-intensities except \(z_t\). For the boundary states \(z_0\) and \(z_{T+1}\), there is no likelihood contribution, so they are updated from their Gaussian full conditionals alone. For \(t=1,\dots,T\), the full conditionals are
\[
p(z_t\mid \bm z_{-t},s_t,y_t,\bm\vartheta)\;\propto\;
\begin{cases}
\mathcal N\!\bigl(z_t;m_t,v_t\bigr), & s_t=0,\\[6pt]
\mathcal N\!\bigl(z_t;m_t,v_t\bigr)\,\mathcal P\!\bigl(y_t;\lambda_t\bigr), & s_t=1,
\end{cases}
\qquad \lambda_t=\exp(z_t),
\]
while for \(t\in\{0,T+1\}\),
\[
p(z_t\mid \bm z_{-t},\bm\vartheta)\;\propto\;\mathcal N\!\bigl(z_t;m_t,v_t\bigr).
\]

The conditional moments follow from standard Gaussian conditioning:
\[
m_t=-\frac{1}{P_{tt}}\sum_{j\neq t} P_{tj} z_j,
\qquad
v_t=\frac{1}{P_{tt}},
\]
where \(P_{tj}\) denotes entry \((t,j)\) of \(\bm P(\bm\vartheta)\). Each univariate full conditional is proper, and we sample \(z_t\) using an adaptive Metropolis--Hastings step \citep{roberts2009examples} that gradually diminishes adaptation of the proposal variance during MCMC and targets an acceptance rate of \(0.234\). The Poisson term \(\mathcal P(y_t;\lambda_t)\) dominates whenever \(y_t\) is large, effectively collapsing the conditional toward a near point mass. In the empirical examples considered below, we therefore did not encounter notable MCMC efficiency problems despite employing single-site updates.

\textbf{ii. Updating $\bm{s}$.} To update the indicators $s_t$, fix $s_t = 1$ for all $t$ where $y_t>0$. For observations where $y_t=0$, sample $s_t$ from the conditional posterior $p(s_t|y_t, z_t, \pi)$. Recall that the prior probability of the sampling path being active is $\pi$ and that $p(y_t=0|z_t) = \mathcal{P}(0; \lambda_t) = \exp(-\lambda_t)$. Using Bayes' theorem, the conditional posterior $p(s_t|y_t, z_t, \pi)$ is hence Bernoulli with 

\begin{equation*}
    p(s_t=1|y_t, z_t, \pi) = \frac{\pi\exp(-\lambda_t)}{(1-\pi)+\pi\exp(-\lambda_t)} \quad \quad \lambda_t = \exp(z_t).
\end{equation*}

\textbf{iii. Updating $\pi$.} The conditional posterior density $p(\pi|\bm{s})$ is Beta $\pi \sim \mathcal{B}(1 + \sum_{t=1}^T s_t, 1+T - \sum_{t=1}^T s_t)$.

\textbf{iv. Updating $\bm{\vartheta}$.} Let $\Delta z_{t} = z_{t} - z_{t-1}$. The hyperparameter updates depend on the choice of the innovation density $f_{\varepsilon}$. We present the posterior updates for four distributional assumptions below.

\paragraph*{Option 1: Gaussian Innovations}

When $\varepsilon_{t} \sim \mathcal{N}(0, \sigma^{2})$, the variance has a conjugate inverse-gamma posterior:

\[
\sigma^{2} \mid \bm{z} \sim \mathcal{IG}\left(
2.5 + \frac{T+1}{2}, 0.5 + \frac{1}{2}\sum_{t=1}^{T+1}(\Delta z_{t})^{2}
\right),
\]

For the latent precision updates, set $k_t = \sigma^{-2}$ for all $t$.

\paragraph*{Option 2: Student-$t$ Innovations}
We employ the scale mixture representation of the Student-$t$ distribution:
\begin{align*}
\Delta z_{t} \mid \omega_{t}, \sigma^{2} &\sim \mathcal{N}\left(0, \frac{\sigma^{2}}{\omega_{t}}\right),\qquad
\omega_{t} \mid \nu \sim \mathcal{G}\left(\frac{\nu}{2}, \frac{\nu}{2}\right),
\end{align*}
with prior $\nu - 3 \sim \text{Exp}(1/6)$ ensuring the first three moments of the innovation density exist. The resulting conditional posteriors are:
\begin{align*}
\sigma^{2} \mid \bm{\omega}, \bm{z} &\sim \mathcal{IG}\left(
2.5 + \frac{T+1}{2}, \;
0.5 + \frac{1}{2}\sum_{t=1}^{T+1} \omega_{t} \Delta z_{t}^{2}
\right), \\[6pt]
\omega_{t} \mid \nu, \sigma^{2}, \bm{z} &\sim \mathcal{G}\left(
\frac{1 + \nu}{2}, \;
\frac{\nu}{2} + \frac{\Delta z_{t}^{2}}{2\sigma^{2}}
\right), \quad t = 1, \ldots, T+1.
\end{align*}

The degrees of freedom $\nu$ are updated via adaptive Metropolis--Hastings steps on the log scale. Propose $\log\nu^{\ast} \sim \mathcal{N}(\log\nu, \tau_{\nu})$ and accept with probability:
\[
\alpha = \min\left\{1, \;
\frac{\nu^{\ast}}{\nu} 
\frac{p(\bm{\omega} \mid \nu^{\ast})}{p(\bm{\omega} \mid \nu)} \frac{p(\nu^*)}{p(\nu)}
\right\},
\]
where the adaptation of $\tau_{\nu}$ targets an acceptance rate of 0.234. For the latent precision updates, set $k_t = \omega_t/\sigma^2$.

\paragraph*{Option 3: Finite Gaussian Scale Mixture}
Consider $H$ mixture components with weights $\bm{\eta} = (\eta_1, \ldots, \eta_H)$, component variances $\sigma^{2}_{h}$, and allocation indicators $\rho_{t} \in \{1, \ldots, H\}$:
\begin{align*}
\Delta z_{t} \mid \rho_{t} = h, \sigma^{2}, \sigma^{2}_{h} &\sim \mathcal{N}(0, \sigma^{2}\sigma^{2}_{h}),\qquad \sigma^{2}_{h} \sim \mathcal{IG}(2.5, 0.5),\\
\rho_{t} \mid \bm{\eta} &\sim \text{Categorical}(\bm{\eta}),\qquad
\bm{\eta} \sim \text{Dirichlet}(1, \ldots, 1).\\
\end{align*}

Define $R_{h} = \sum_{t=1}^{T+1} \mathbbm{1}(\rho_{t} = h)$ as the number of allocations to component $h$. The resulting conditional posteriors are:
\begin{align*}
\sigma^{2} \mid \sigma^2_{\rho_t}, \bm{z} &\sim \mathcal{IG}\left(
2.5 + \frac{T+1}{2}, \;
0.5 + \frac{1}{2}\sum_{t=1}^{T+1} \sigma^{-2}_{\rho_{t}} \Delta z_{t}^{2}
\right), \\[6pt]
\Pr(\rho_{t} = h \mid \cdot) &\propto \eta_{h} \cdot \varphi(\Delta z_{t}; 0, \sigma^{2}\sigma^{2}_{h}), \\[6pt]
\bm{\eta} \mid \bm{\rho} &\sim \text{Dirichlet}(1 + R_{1}, \ldots, 1 + R_{H}), \\[6pt]
\sigma^{2}_{h} \mid \cdot &\sim \mathcal{IG}\left(
2.5 + \frac{R_{h}}{2}, \;
0.5 + \frac{1}{2\sigma^{2}} \sum_{t: \rho_{t} = h} \Delta z_{t}^{2}
\right),
\end{align*}
where $\varphi(\cdot; \mu, \sigma^2)$ denotes the Gaussian probability density function with mean $\mu$ and variance $\sigma^2$. For latent precision updates, set $k_t = (\sigma^{2}\sigma^{2}_{\rho_t})^{-1}$. We use $H=2$ mixture components in the applications below.

\paragraph*{Option 4: Stochastic volatility}
For the SV case we employ the auxiliary-mixture sampler of \citet{kim1998stochastic},
combined with the multivariate Gaussian joint density representation for state-space formulations (\citealp{mccausland2011simulation});
see \citet{kastner2016dealing} and the \texttt{stochvol} \textsf{R} package for an
efficient software implementation. The estimated variance pathways $\sigma^2_t = \exp(h_t)$ can then be used to update the latent precisions by setting $k_t = \sigma^{-2}_t$.

\subsection{Forecasting and Log Predictive Score Evaluation}

To sample from the posterior predictive distribution of \(y_{T+1}\), for each MCMC draw
\(m\) sample \(s_{T+1}^{(m)}\sim\mathrm{Bernoulli}(\pi^{(m)})\). If \(s_{T+1}^{(m)}=0\),
set \(y_{T+1}^{(m)}=0\). If \(s_{T+1}^{(m)}=1\), sample \(z_{T+1}\) from the state
evolution, which under the random-walk prior satisfies
\(z_{T+1}\mid z_T,k_{T+1}\sim\mathcal{N}(z_T,k_{T+1}^{-1})\), and then sample
\(y_{T+1}^{(m)}\mid z_{T+1}^{(m)}\sim\mathcal{P}(\exp(z_{T+1}^{(m)}))\).
Forecasts conditional on \(s_{T+1}=1\) are obtained by skipping the Bernoulli step.

For log predictive distribution evaluation we integrate out \(z_{T+1}\) numerically. For each
posterior draw \(m\), define
\[
g^{(m)}(y)
= \int \mathcal{P}\!\bigl(y;\exp(z)\bigr)\, p\!\bigl(z_{T+1}=z \mid z_T^{(m)},k_{T+1}^{(m)}\bigr)\,dz,
\]
which we approximate using Gauss--Hermite quadrature. The corresponding full
zero-inflated predictive pmf is
\[
p^{(m)}(y\mid \bm y)=
\begin{cases}
(1-\pi^{(m)})+\pi^{(m)}g^{(m)}(0), & y=0,\\[4pt]
\pi^{(m)}g^{(m)}(y), & y\ge 1.
\end{cases}
\]
Given \(M\) posterior draws, the estimated log predictive distribution at \(y_{T+1}\) is
\[
\widehat{\log p}(y_{T+1}\mid\bm{y})
= \log\biggl\{\frac{1}{M}\sum_{m=1}^{M} p^{(m)}(y_{T+1}\mid\bm{y})\biggr\},
\]
while the score conditional on \(s_{T+1}=1\) is
\[
\widehat{\log p}(y_{T+1}\mid s_{T+1}=1,\bm{y})
= \log\biggl\{\frac{1}{M}\sum_{m=1}^{M} g^{(m)}(y_{T+1})\biggr\}.
\]
\subsection{Simulation Exercises}

We evaluate the proposed mechanisms and posterior simulation algorithms in a controlled setting using a simulation study. Full details are provided in Supplementary Section~\ref{app:sim}; here we summarize only the main insights for brevity. We compare zero-inflated homoskedastic and stochastic volatility Poisson random walk models against a naive homoskedastic Poisson approach that treats all zeros as arising from the sampling count process. In the first setting, the random walk innovation variance is simulated as constant. The homoskedastic Poisson random walk recovers parameters well, while the stochastic volatility model approximates the constant variance accurately, though, as expected, with somewhat lower efficiency. The naive model severely overestimates variance by misattributing occasional drops to zero as part of the sampling process. In a second setting, the innovation variance is assumed to oscillate between low and high volatility periods. The stochastic volatility model accurately recovers the underlying volatility dynamics. The zero-inflated homoskedastic model produces averaged estimates that overestimate volatility during tranquil periods and underestimate it during volatile periods. The naive Poisson random walk performs worst, severely overestimating innovation variance by ignoring zero inflation.

Overall, the stochastic volatility specification proves sufficiently flexible to accommodate both constant and time-varying volatility regimes, nesting the simpler homoskedastic model as a special case. This flexibility translates to strong empirical performance in the application presented in Section~\ref{sec:application}. Regarding zero inflation, we find no substantial differences between the stochastic volatility and homoskedastic Poisson specification in the settings considered. While the stochastic volatility model performs similarly or marginally better in classifying zeros as structural, both models classify truly structural zeros with very high accuracy.

\section{Modeling Irregular Migration in the Mediterranean Sea and the English Channel}
\label{sec:application}

\subsection{Predictive Performance and Benchmarking}

Before discussing inferential results under any of the proposed specifications, we first assess distributional forecasting performance under alternative choices of $f_{\varepsilon}$. The purpose of this exercise is not to claim that any single specification dominates all conceivable competitors, but rather to identify which model components (e.g., heavy tails or stochastic volatility) are required for reasonable distributional forecasting performance in the context of irregular maritime migration.

We conduct a benchmarking exercise in which we repeatedly split the data into a training sample of length $T-250$ and a one-step-ahead hold-out observation, whose distribution we aim to predict. We repeat this procedure for 250 hold-out periods using a rolling-window scheme. For each model, we collect 75,000 posterior samples after a burn-in period of 7,500 iterations. We then compute several performance metrics, including cumulative log-predictive scores, root mean square error, and correlations between posterior mean forecasts and realized values. To evaluate tail calibration, we additionally report empirical coverage with respect to the predictive tail quantiles $q_{01}$, $q_{05}$, $q_{10}$, $q_{90}$, $q_{95}$, and $q_{99}$.

\subsubsection{Operational Target and the Role of Structural Zeros} 

As our ultimate aim is to provide operationalizable forecasts \textit{conditional on the sampling regime being active} and to accurately capture the upper tail of the predictive distribution, we report baseline results based on predictive distributions conditional on $s_{T+1}=1$ and compute out-of-sample metrics using only observations with $y_t>0$, thereby focusing evaluation on weeks with realized crossing activity. This evaluates how well a model predicts the magnitude of arrivals when they occur. In the context of policy preparedness, this conditional forecasting target is arguably the most operationally relevant. For completeness, and as an additional robustness check, we also report results that integrate over uncertainty in $s_{T+1}$ and include all hold-out observations in the supplementary material; these are referenced and briefly discussed below.

\begin{table}[t]
  \centering
  \caption{Results of the Predictive Exercise and Hold-out Evaluation.}
  \label{tab:fc_res_cond}
  \adjustbox{width=\textwidth}{%
  \begin{threeparttable}
    \begin{tabular}{llrrrrrrrrrr}
      \toprule
      \multicolumn{2}{c}{} &
      \multicolumn{1}{c}{\textbf{LPS}} &
      \multicolumn{1}{c}{\textbf{RMSE}} &
      \multicolumn{1}{c}{\textbf{Corr.}} &
      \multicolumn{1}{c}{\textbf{$q_{01}$}} &
      \multicolumn{1}{c}{\textbf{$q_{05}$}} &
      \multicolumn{1}{c}{\textbf{$q_{10}$}} &
      \multicolumn{1}{c}{\textbf{$q_{90}$}} &
      \multicolumn{1}{c}{\textbf{$q_{95}$}} &
      \multicolumn{1}{c}{\textbf{$q_{99}$}} &
      \multicolumn{1}{c}{\textbf{n}}\\
      \midrule
ITA & Gaussian (excl. zeros) & -2144.436 & 7.563 & 0.366 & 0.020 & 0.057 & 0.081 & 0.899 & 0.931 & 0.980 & 247\\
ITA & Gaussian (incl. zeros) & -2180.422 & 7.564 & 0.367 & 0.004 & 0.016 & 0.040 & 0.931 & 0.955 & 0.984 & 247\\
ITA & Gaussian & -2144.371 & 7.565 & 0.365 & 0.020 & 0.057 & 0.081 & 0.899 & 0.931 & 0.980 & 247\\
ITA & Mixture & -2135.893 & 7.565 & 0.364 & 0.012 & 0.057 & 0.081 & 0.895 & 0.935 & 0.984 & 247\\
ITA & Student's t & -2138.267 & 7.565 & 0.366 & 0.012 & 0.057 & 0.117 & 0.895 & 0.931 & 0.984 & 247\\
ITA & Stoch. Vol & -2121.401 & 7.562 & 0.366 & 0.008 & 0.057 & 0.101 & 0.907 & 0.955 & 0.992 & 247\\
\addlinespace
UK & Gaussian (excl. zeros) & -1726.586 & 6.546 & 0.367 & 0.023 & 0.068 & 0.108 & 0.869 & 0.919 & 0.973 & 222\\
UK & Gaussian (incl. zeros) & -1854.025 & 6.594 & 0.335 & 0.000 & 0.000 & 0.005 & 0.883 & 0.905 & 0.955 & 222\\
UK & Gaussian & -1726.057 & 6.545 & 0.367 & 0.023 & 0.068 & 0.108 & 0.869 & 0.914 & 0.973 & 222\\
UK & Mixture & -1724.793 & 6.545 & 0.367 & 0.023 & 0.068 & 0.108 & 0.869 & 0.919 & 0.973 & 222\\
UK & Student's t & -1725.238 & 6.546 & 0.366 & 0.014 & 0.072 & 0.113 & 0.860 & 0.914 & 0.977 & 222\\
UK & Stoch. Vol & -1721.395 & 6.547 & 0.368 & 0.009 & 0.059 & 0.117 & 0.878 & 0.932 & 0.986 & 222\\
      \bottomrule
    \end{tabular}
    \begin{tablenotes}
      \footnotesize
      \item Note: Columns report the cumulative log predictive score (LPS), root mean square error (RMSE) on the logarithmic scale, correlation of forecasts and true values (Corr.), and empirical coverage with respect to the predictive 1st, 5th, 10th, 90th, 95th, and 99th quantiles ($q_{01}-q_{99}$), with $n$ denoting the number of non-zero hold-out observations. Results are summaries across out-of-sample periods where $y_{T+1}>0$ and conditional on $s_{T+1}=1$. Second column refers to the innovation model for the Poisson random walk. `Gaussian (excl. zeros)' is a homoskedastic model where all zeros are treated as missing. `Gaussian (incl. zeros)' is a homoskedastic model where all zeros are considered part of the sampling process. All other models use an explicit zero inflation mechanism to estimate $s_t$.
    \end{tablenotes}
  \end{threeparttable}
}
\end{table}

\subsubsection{Baseline Predictive Results}

Tab.~\ref{tab:fc_res_cond} reports the baseline out-of-sample results. They confirm our initial assessment that accurate point predictions for irregular sea border crossings are difficult to obtain. The generally low correlation between point forecasts and realized values highlights the limited point predictability of the irregular maritime border-crossing process, reinforcing our decision to emphasize distributional forecasts. For the Italian data, specifications with heavy-tailed innovations -- implemented either via mixture models or $t$-distributions -- outperform the homoskedastic Gaussian benchmark in terms of log predictive scores and tail coverage.

\begin{figure}[t]
     \centering
     \begin{subfigure}[b]{0.49\linewidth}
         \centering
         \includegraphics[width=\linewidth]{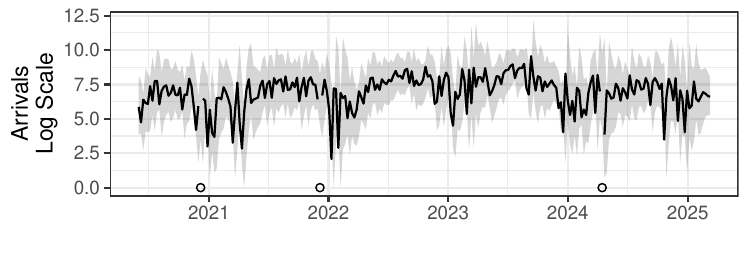}
     \end{subfigure}
          \begin{subfigure}[b]{0.49\linewidth}
         \centering
         \includegraphics[width=\linewidth]{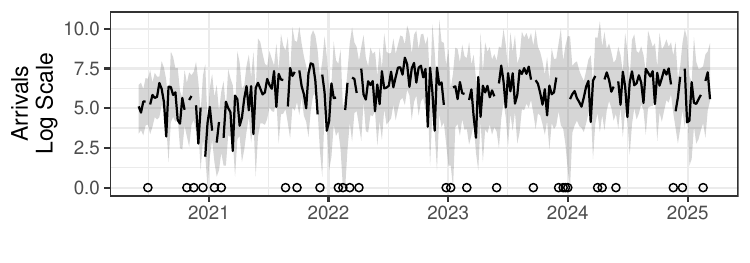}
     \end{subfigure}\\
     \begin{subfigure}[b]{0.49\linewidth}
         \centering
         \includegraphics[width=\linewidth]{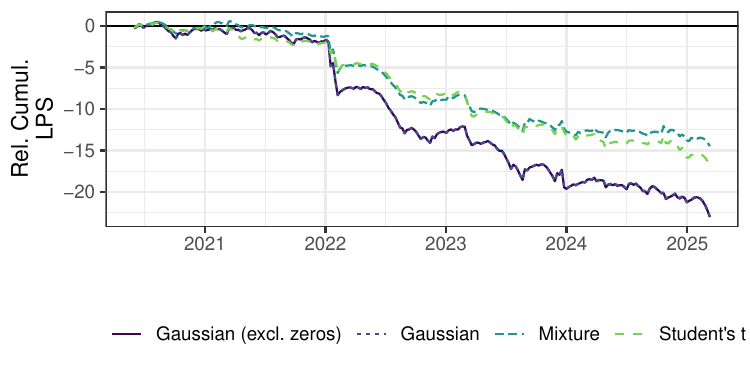}
     \end{subfigure}
     \begin{subfigure}[b]{0.49\linewidth}
         \centering
         \includegraphics[width=\linewidth]{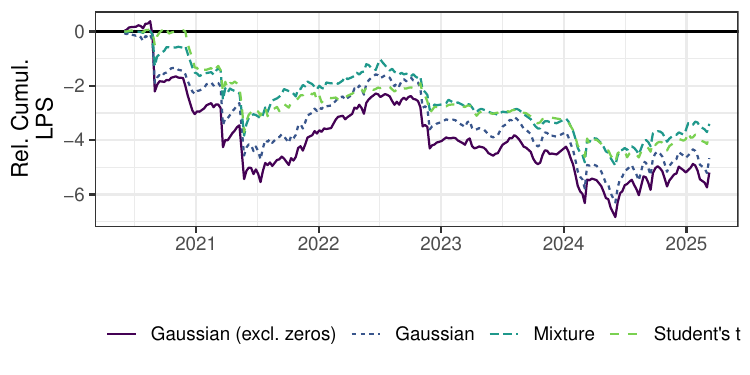}
     \end{subfigure}
     
        \caption{Top row: One-step-ahead predictive distributions and realized observations (black line) over the 250 hold-out periods. Zero observations are highlighted separately on the $x$-axis.  Grey shaded areas are 90\% credible intervals.  Bottom row: Cumulative log predictive scores relative to the stochastic volatility model. The Poisson model without zero inflation is not shown due to non-competitive performance. Left column: Mediterranean crossing data. Right column: English Channel crossing data. Hold-out period runs from 2020W23 to 2025W11. Results are computed across out-of-sample periods where $y_{T+1}>0$ and conditional on $s_{T+1}=1$.}
        \label{fig:cumul-lps}
\end{figure}

The strongest performance in both case studies is achieved by specifications that incorporate stochastic volatility, providing compelling evidence of \textit{persistent} time variation in volatility in the underlying migration intensity process. This suggests that the process alternates between relatively stable periods and episodes of heightened fluctuation. Fig.~\ref{fig:cumul-lps} further illustrates this point by tracking cumulative log predictive scores relative to the stochastic volatility model over the hold-out period. For the Italian data, the gains from stochastic volatility are consistent throughout, and are particularly pronounced during high-volatility episodes (e.g., early 2022), when the stochastic volatility model is able to adapt more flexibly. For the UK data, the gains of stochastic volatility relative to static heavy-tailed specifications are weaker and less uniform over time. Nonetheless, over the considered hold-out period the improvements remain substantial, and we view stochastic volatility as a strong benchmark also for this case study. Future work may further validate or revise this conclusion as longer time series become available.

Turning to tail calibration, empirical tail coverage closely tracks nominal coverage under the stochastic volatility model. It is noteworthy that a highly parsimonious specification achieves good tail calibration for a highly complex and volatile social process. Accurate tail quantile estimates and forecasts are particularly useful in this context, as they can be used for contingency planning and as operational risk indicators. From the perspective of statistical decision theory, an important consideration is that quantiles solve optimal decision problems under (asymmetric) linear loss functions. If such loss functions could be elicited from forecast users (e.g., migration practitioners or policymakers), even approximately, policy and operational preparedness could be substantially improved based on distributional forecasts \citep[see e.g.][]{bijak2010}. So far, such elicitation has not been carried out in the context of migration, although especially given the high policy salience, there are strong arguments for implementing it as a part of broader preparedness efforts \citep{ec2020}. Additionally, an explicit elicitation of potential impacts of different migration scenarios would be in line with existing examples of preparedness and contingency planning \cite[e.g. \emph{National Risk Register},][]{riskreg25}, and recent work on infectious disease response \citep{mills2025},  which use probabilistic assessment for risk management purposes. 

In addition, quantiles are invariant under monotonic transformations. This implies that if the policy objective functions of different decision makers (such as the cost of running reception centers or time of processing asylum applications) can be expressed as monotonic transformations $g(\cdot)$ of the number of irregular migrants $y_t$, then under linear loss functions, the solutions to such decision problems are simply the values of $g(\cdot)$ evaluated at the respective quantiles of the predictive distribution of $y_t$. This feature additionally underscores the versatility of the approach, where different users, with different objectives and loss functions, could use the same underlying probability distributions to support their decisions, as long as these distributions are well-calibrated, including in the tails.

Finally, quantile estimates and forecasts also admit straightforward operational interpretations in our context. For example, the 90th percentile corresponds to an approximate 10\% weekly probability of exceeding the indicated threshold, implying an expected exceedance about once every 10 weeks (roughly every 2--3 months). Likewise, the 95th percentile corresponds to about a 5\% weekly probability, or one exceedance approximately every 20 weeks (around every 5 months). The 99th percentile implies roughly a 1\% weekly probability, corresponding to an extreme event about once every 100 weeks, i.e., close to once every two years. While these interpretations rely on an independence approximation and should therefore be viewed as heuristic, they nonetheless yield transparent and interpretable indicators that can serve as starting points for contingency planning. Developing more comprehensive operational risk indicators for this domain remains an avenue for future research.

\subsubsection{Further Predictive Results and Insights}

Tab.~\ref{tab:fc_res_all} in the supplementary material reports log predictive scores and tail coverage when marginalizing over $s_t$ and evaluating performance on all hold-out observations (including weeks with $y_t=0$). Supplementary Fig.~\ref{fig:cumul-lps-full} presents the corresponding posterior predictive distributions and cumulative log predictive score plots. Overall, these results lead to very similar conclusions, but indicate slightly worse empirical coverage in the lower tail.

This difference primarily reflects our convenient, but simplifying, assumption of a constant probability $\pi$ over the sample. A constant $\pi$ has two implications for forecasting. First, when the estimated probability of a structural zero $(1-\pi)$ is sufficiently large, the lowest predictive quantiles become exactly zero (compare Fig.~\ref{fig:cumul-lps-full}). This also affects in-sample fits; see Supplementary Fig.~\ref{fig:both-fits}. The effect is modest for the Italian series, where zeros are rare and $\pi$ is estimated close to one. The quantiles are therefore similar under conditioning on $s_t=1$ and marginalizing over $s_t$. In contrast, for the UK series with a larger fraction of zeros, the differences are more substantial. For example, $q_{05}$ is equal to zero throughout the sample period when marginalizing over $s_t$. Second, a constant $\pi$ typically overstates the prior probability of structural zeros during periods in which they are uncommon and vice versa. In our application this does not appear to materially affect overall predictive performance, as the likelihood is typically informative about the separation between structural zeros and non-zero observations. Nevertheless, for forecasting it implies that the predictive distribution may be biased, with size and direction of the bias depending on the period under consideration. Model extensions and generalizations that relax the constant-$\pi$ assumption are discussed in Sec.~\ref{sec:conclusion}.

\subsection{Model Fit, Volatility Estimates, Zero Inflation Patterns}

\begin{figure}[t]
     \centering
     \begin{subfigure}[b]{0.49\linewidth}
         \centering
         \includegraphics[width=\linewidth]{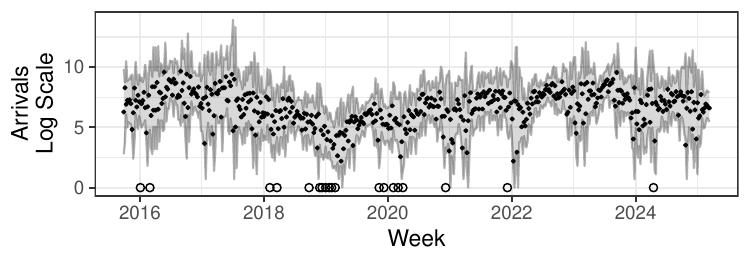}
     \end{subfigure}
          \begin{subfigure}[b]{0.49\linewidth}
         \centering
         \includegraphics[width=\linewidth]{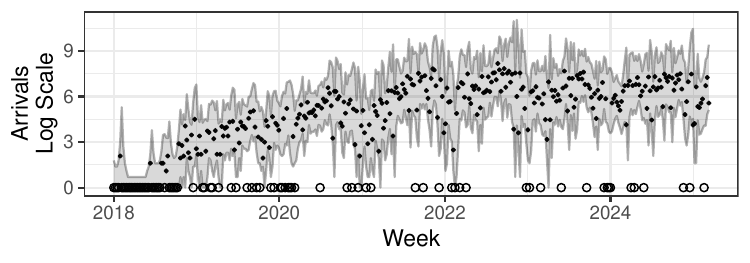}
     \end{subfigure}\\
     \begin{subfigure}[b]{0.49\linewidth}
         \centering
         \includegraphics[width=\linewidth]{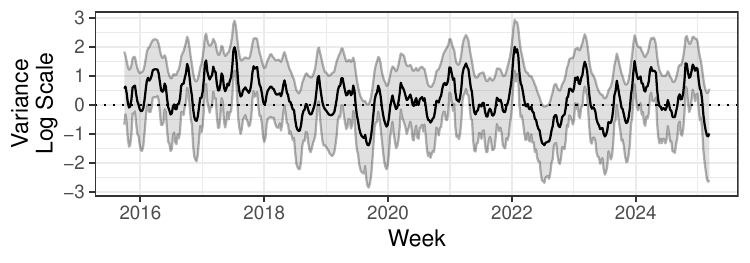}
     \end{subfigure}
     \begin{subfigure}[b]{0.49\linewidth}
         \centering
         \includegraphics[width=\linewidth]{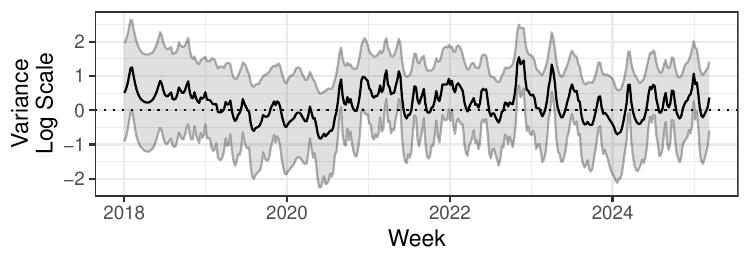}
     \end{subfigure}\\
      \begin{subfigure}[b]{0.49\linewidth}
         \centering
         \includegraphics[width=\linewidth]{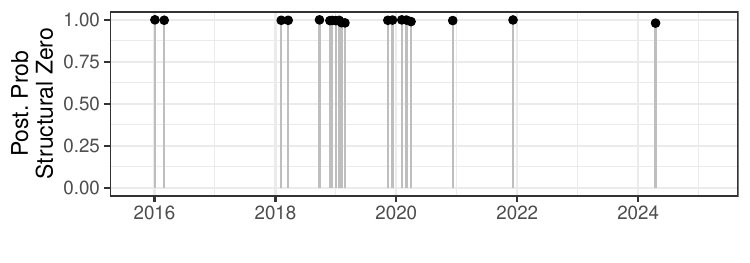}
     \end{subfigure}
     \begin{subfigure}[b]{0.49\linewidth}
         \centering
         \includegraphics[width=\linewidth]{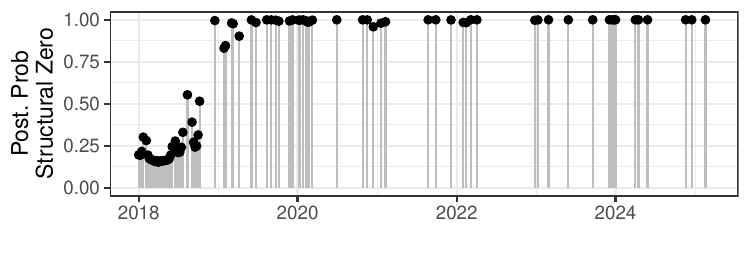}
     \end{subfigure}
     
        \caption{Top row: Crossing counts and distributional fit by sample week on a logarithmic scale. Zero counts are highlighted separately on the zero line. Middle row: Estimated posterior density of $h_t$. Bottom row: Estimated posterior mean probability of $y_t$ being a structural zero. Left column: Mediterranean crossings by sample week from 2015W40 to 2025W11. Right column: English Channel crossings by sample week from 2018W01 to 2025W11. Solid lines are posterior means. Grey shaded areas indicate 90\% credible intervals.}
        \label{fig:model_fit}
\end{figure}

Fig.~\ref{fig:model_fit} displays \emph{smoothed one-step-ahead predictive distributions} for the count data models. At each time point $t$, these distributions are obtained by propagating draws from the full-sample (smoothed) posterior of the latent intensity one step forward through the state-space model and then generating the corresponding count outcome. Repeating this procedure yields a Monte Carlo approximation to the model-implied distribution of $y_t$ conditional on $s_t=1$ and on the full information in $\bm y$.\footnote{We use this quantity as an illustrative in-sample distributional fit diagnostic. It differs from the real-time one-step-ahead predictive distributions in Fig.~\ref{fig:cumul-lps}, which condition only on information available up to $t-1$.} The middle row of Fig.~\ref{fig:model_fit} shows posterior density estimates of the log variances $h_t=\log(\sigma_t^2)$, illustrating the time-varying uncertainty captured by the stochastic volatility model. All reported summaries are based on 75{,}000 posterior draws after a burn-in of 7{,}500 iterations.

For both the UK and Mediterranean crossings data, Fig.~\ref{fig:model_fit} helps explain why the stochastic volatility specification outperforms competing models in predictive terms. In the Mediterranean case, the estimated variance exhibits clear alternation between high-volatility episodes (e.g., early 2022) and more tranquil periods (e.g., mid-2022). Similar, albeit less pronounced, patterns emerge in the UK case study, consistent with the smaller performance gap between the stochastic volatility model and the competing specifications. In both applications, the stochastic volatility component captures shifts between tranquil and volatile regimes, allowing the predictive distribution to widen or tighten adaptively and thereby improving predictive distribution estimates.

In terms of the zero-inflation component, the bottom row of Fig.~\ref{fig:model_fit} reports the posterior probabilities that an observed zero corresponds to a structural zero. For the Mediterranean data, the posterior mean $\hat{\pi}=0.96$ indicates that the sampling regime is active most of the time. When zeros do occur, the model assigns them to the structural-zero state ($s_t=0$) with near-unit probability. As a result, the implied zero-inflation probability closely matches the empirical frequency of zeros in the data (3.8\%). For the UK data, we obtain $\hat{\pi}=0.85$, compared to an empirical zero frequency of 22.3\%, indicating that not all zeros are classified as structural. In particular, most zeros prior to 2019 are attributed to the sampling process ($s_t=1$), whereas zeros after 2019 are predominantly classified as structural ($s_t=0$). This pattern reflects that migration intensity is comparatively low early in the sample. Zeros in that period are interpreted as arising from this low intensity (i.e., few or no attempts), rather than from temporary conditions that render crossings effectively impossible.

\subsection{Further Exploratory Exercises} 
\label{sec:explorative}

To gain deeper insight into the drivers of time-varying volatility patterns and structural zeros in sea migration data, we conduct additional empirical analyses. Specifically, we aim to explain the posterior mean of (i) log variances and (ii) the logit-transformed \emph{a posteriori} structural-zero probabilities for each $t$ in both case studies using linear regression. For completeness, we also consider (iii) an exploratory linear regression model with the logarithm of weekly border crossings as the outcome variable.\footnote{Note that this procedure is of course quite imperfect and exploratory in nature, given that it ignores various aspects such as correlations across time or uncertainty in the latent outcome variables, among others. Nevertheless, it provides some first-order insights into process dynamics and potential drivers.} As explanatory variables, these models include monthly and annual effects to capture seasonal patterns and long-term trends as well as potential policy shifts. Additionally, we incorporate a variable measuring wave height in the broader crossing regions (see Supplementary Fig.~\ref{fig:map}) to approximate prevailing weather and sea conditions.\footnote{Data on significant wave height are collected by the European Centre for Medium-Range Weather Forecasts (ECMWF) and are available at \url{https://cds.climate.copernicus.eu/datasets/reanalysis-era5-single-levels} (as of 7 August 2025).}

The results of these exploratory models, which are detailed in Supplementary Tab.~\ref{tab:regression-results}, reveal several distinct dynamics. We find that both seasonal and annual patterns predict variance patterns in the Mediterranean case study, with the summer months typically being the most tranquil. A similar pattern is visible in the English Channel example, although it is less pronounced. In both case studies, the wave height variable is positively related to log variance, though the association is relatively weak once seasonality is controlled for in the same model. This likely reflects that wave height and broader meteorological conditions are themselves strongly seasonal.

For both case studies, structural zeros do not exhibit clear seasonal patterns. Wave height, however, is a strong predictor of structural zeros in both settings, underscoring the relevance of weather and sea conditions in maritime migration; compare also the findings in \citet{wood2025factors}. Given the small number of zero weeks in the Italian data, we find no strong annual effects, whereas for the UK the later years show an increase in structural zeros. Note that structural zeros may reflect interception by coast guards, weather effects, accidents, or lack of smuggler availability. We can only explore one of these mechanisms explicitly, since reliable data on interceptions or smuggler activity are typically not publicly available. Finally, wave height also has strong effects on overall arrival intensity on the log scale in both case studies, suggesting increased accident risk and a greater reluctance to initiate a journey in adverse weather conditions as potential underlying channels.

Overall, the explanatory power of the covariates -- as measured by the coefficient of determination -- remains modest, suggesting that a broader constellation of factors likely drives the volatility, feasibility, and magnitude of crossings. Our findings therefore indicate that at least half of the variation in volatility and feasibility cannot be attributed to long-run trends, annual effects, seasonal patterns, or weather and maritime conditions. This substantial unexplained component likely reflects unmeasured factors such as fluctuations in smuggling operations, coast guard activities, and upstream determinants of the potential migrant population (e.g., conflict events and political instability). Importantly, both the reported $R^2$ values and the covariate importance assessments should be interpreted cautiously. They are overly confident, as the regression models do not account for uncertainty in the estimated volatility paths and in the structural-zero probabilities, and due to the presence of autocorrelation.

Future work could incorporate covariate-driven volatility or structural-zero dynamics directly into the model to strengthen latent volatility estimation, rather than using covariates only in post-hoc exploratory regressions. This would directly connect to related zero-inflated models to analyze migration dynamics (\citealp{wood2025factors}; \citealp{zhang2026spatiotemporal}). For real-world deployment, however, such an approach would face practical challenges, since weather data and other relevant covariates often become available only ex post and would therefore also need to be forecast, introducing additional modeling complexity.

\section{Discussion and Concluding Remarks}
\label{sec:conclusion}

\subsection{Policy Implications}

The empirical evidence in favor of a random walk with persistent time-varying volatility in border crossing activity has at least three direct policy implications. First, our finding that irregular migration is well approximated by a random walk confirms earlier intuitions about the inherent difficulty of predicting these flows \citep{bijak2010}. Because random walks imply limited forecastability and uncertainty that expands rapidly with the forecast horizon, turning points may easily be missed, even by elaborate point forecasting tools. Policymakers should therefore avoid relying on medium-run forecasts and rigid annual quotas in reception infrastructure. Instead, the focus should be on real-time monitoring tools and on maintaining flexible surge capacity (e.g., modular reception centers, standby search-and-rescue assets, or dynamic solidarity mechanisms) that can scale up or down as the migration process unfolds.

Second, the evidence for persistence in volatility patterns suggests clustering of calm and volatile periods, with long stretches of low variance alternating with bursts of high variance. This mirrors field observations that Mediterranean crossings can remain subdued for months and then increase sharply within weeks when, for example, a civil-war front shifts, a smuggling route re-opens, or the business model of smuggling networks changes. From a policy perspective, this points to treating migration management as a \textit{risk management} problem, not only as flow management. Risk management explicitly requires taking into account both the \textit{uncertainty} and potential \textit{impact} of unforeseen events (e.g. \citeauthor{riskreg25}, \citeyear{riskreg25}; for a migration example, see \citeauthor{bijaketal2019}, \citeyear{bijaketal2019}). Our results suggest that an important part of the uncertainty considerations may be related to the \textit{volatility} of flows. During low-volatility windows, policymakers may invest in contingency planning and insurance-type funding (e.g., an EU-wide migration contingency fund), so that cash, assets, and staff can be pre-positioned for the next high-volatility spell and scaled up rapidly as demand increases.

Third, a random walk with stochastic volatility implies a heavy-tailed distribution of outcomes, in the sense that extreme events occur more frequently than under standard homoskedastic models. Conceptually, occasional extremes should therefore not be treated as ``unexpected outliers'' or ``black swan'' events, but rather as a structural feature of irregular migration processes. Policy should accordingly prioritize systems that are robust to tail risk. Examples include contingency plans that ensure sufficient search-and-rescue capacity for ``worst-case'' scenarios, fast-track legal pathways that can relieve pressure once a surge begins, the negotiation of standby humanitarian corridors with coastal states, or trigger-based burden-sharing clauses to reduce pressure on countries of first arrival (e.g., automatic relocation once arrivals exceed a pre-defined high percentile).

Note that replacing our simpler unit-root assumption with a less stringent but still highly persistent mean-reverting process (e.g., an AR(1) model with persistence parameter close to one) would only slightly alter these conclusions. However, an AR(1) regime would suggest two additional policy levers. First, interventions that effectively reduce persistence -- for instance through diplomacy in origin states, legal work-migration channels, or humanitarian cash transfers -- could speed up the decay of shocks and spikes. Second, because the resulting predictive intervals are large but not unbounded, risk-based insurance and reinsurance instruments (e.g., bonds that pay out when arrivals exceed a high percentile) may be priced more realistically, especially when coupled with statistical decision theory. In short, high persistence implies that shocks linger, yet even weak mean reversion leaves scope for proactive measures that accelerate a return toward long-run ``normality''.

In summary, viewing irregular maritime migration through the lens of a random walk with stochastic volatility shifts the policy focus from forecasting and flow management toward risk management and preparedness, and from average-case planning toward governance informed by volatility and tail risk. While a fully fledged real-world operationalization of these ideas is beyond the scope of this article, the well-calibrated distributional forecasts produced by our model may provide a useful starting point in that direction.

\subsection{Model Extensions and Future Research}

Several avenues for future research remain. First, alternative and more flexible specifications for $f_{\varepsilon}$ could be considered, including asymmetric innovations or combinations of stochastic volatility and heavy tails. Second, a dynamic specification for the sampling probability $\pi_t$ -- for example via dynamic probit models -- would be a natural extension and could improve forecasting performance. Our simplifying assumption of constant $\pi$ is convenient, but does not reflect real-world observations such as the clustering of zeros over time. Moreover, both the $\pi_t$ process and the volatility process could be allowed to co-move with the latent count intensity, which may further strengthen out-of-sample predictive performance, particularly for the zero-inflation component. Finally, the conditional mean dynamics could be explored further by assessing whether a highly persistent, but mean-reverting, specification outperforms the RW(1) benchmark. Overall, in the irregular sea border crossings applications studied here, several modeling choices could be refined to improve log predictive scores or achieve more accurate tail coverage. Nevertheless, based on our empirical exercises, we are confident that a random walk with persistent, time-varying volatility provides a useful, simple, and well-performing benchmark framework for irregular migration processes.

Moreover, our analysis has focused on a univariate setting, but extensions to multivariate count data frameworks could be highly valuable. Potential applications in migration and border crossing data are numerous. For example, jointly modeling crossings at different European entry points could enable borrowing strength across borders, border types, and origin countries when modeling sparse and noisy migration time series. This may also help to provide new, granular evidence on the impact of differentially restrictive visa and migration enforcement regimes on changes in migration routes and preferred destinations, and on switching from legal to illegal modes of crossing borders \citep{czaikahobolth2016}. Conceptually, we expect that a straightforward multivariate extension of our current model could provide a useful baseline for multivariate irregular migration dynamics. At the same time, many applications likely exhibit complex cross-series dependency structures, as we have observed, for instance, in ongoing work on probabilistic modeling of asylum dynamics in the EU-27.

On a final note, ethical considerations are paramount when developing models in this domain. Any predictive approach can be problematic given the risk of dual use or misuse -- for instance, being used to advocate more restrictive border-closure policies or repressive measures, or being exploited by smuggling networks to inform their operations. In our setting, this risk is relatively limited because we model weekly arrivals at the country level, rather than, for example, daily route-specific departures. Moreover, as discussed above, border crossing series are barely point-predictable, which further constrains operational misuse. Ultimately, these concerns intersect with political choices that reflect societal values and are negotiated through democratic processes. At the same time, they must be weighed against the potential benefits of improved preparedness and more effective humanitarian resource allocation, particularly for sea border crossings. Risks of misuse can be further mitigated through ethical oversight, strict data protection protocols, and inclusive stakeholder engagement to ensure that such models serve public safety and humanitarian objectives without undermining individual rights.

%%%%%%%%%%%%%%%%%%%%%%%%%%%%%%%%%%%%%%%%%%%%%%
%% Support information, if any,             %%
%% should be provided in the                %%
%% Acknowledgements section.                %%
%%%%%%%%%%%%%%%%%%%%%%%%%%%%%%%%%%%%%%%%%%%%%%
\begin{acks}[Acknowledgments]

The authors gratefully acknowledge Maximilian Böck for insightful discussions that helped shape the early development of this work. We further thank the associate editor and three anonymous referees for constructive feedback, and the participants of the \textit{Wittgenstein Centre Conference 2025: Demographic Perspectives on Migration in the 21st Century} for helpful comments.

\end{acks}

\begin{acks}[Significance Statement]

Irregular maritime migration -- or "small boat" crossings -- poses major humanitarian and policy challenges, yet statistical understanding of these flows remains limited. We develop a probabilistic model that captures three key features of weekly sea arrival data: persistent migration intensity, periods of calm alternating with volatile times, and occasional weeks with seemingly unexpected zero crossings. Applied to Mediterranean and English Channel data, we find strong evidence that volatility itself changes over time in persistent patterns. Our model produces well-calibrated forecasts, including accurate predictions of rare tail events. This offers a foundation for policy approaches centered on risk management and preparedness rather than point predictions. The methodology extends to other non-stationary count time series with zero inflation.
    
\end{acks}

%%%%%%%%%%%%%%%%%%%%%%%%%%%%%%%%%%%%%%%%%%%%%%
%% Funding information, if any,             %%
%% should be provided in the                %%
%% funding section.                         %%
%%%%%%%%%%%%%%%%%%%%%%%%%%%%%%%%%%%%%%%%%%%%%%
\begin{funding}
Gregor Zens gratefully acknowledges financial support from the European Research Council (ERC) through the \textit{2C-RISK} grant (Grant Agreement No. 101162653). Jakub Bijak's work was supported by the Leverhulme Trust (Grant RC-2018-003) for the Leverhulme Centre for Demographic Science, based on earlier work from the Horizon 2020 project \textit{QuantMig} funded by the European Commission (Grant No. 870299). Views and opinions expressed are, however, those of the authors only and do not necessarily reflect those of the European Union, the European Research Council Executive Agency or the Leverhulme Trust. Neither the European Union nor the granting authorities can be held responsible for them.
\end{funding}

\bibliographystyle{imsart-nameyear} % Style BST file
\bibliography{lit}       % Bibliography file (usually '*.bib')

%%%%%%%%%%% SUPPLEMENT

\newpage
\setcounter{page}{1}
\setcounter{figure}{0}
\setcounter{section}{0}
\setcounter{table}{0}
\renewcommand\thesection{S\arabic{section}}
\renewcommand\theequation{S\arabic{equation}}
\renewcommand\thefigure{S\arabic{figure}}
\renewcommand\thetable{S\arabic{table}}

\begin{center}
\vspace*{2em}
{\Large\textbf{Supplementary Material for ``Dynamic Count Models with Flexible Innovation Processes for Irregular Maritime Migration''}}\\[1.5em]
{\large Gregor Zens$^{1}$ and Jakub Bijak$^{2}$}\\[1em]
{\normalsize $^{1}$Population and Just Societies Program, International Institute for Applied Systems Analysis (IIASA)\\[0.5em]
$^{2}$Leverhulme Centre for Demographic Science, Nuffield Department of Population Health\\and Reuben College, University of Oxford}\\[2em]
\end{center}

\vspace{2em}

\section{Simulation Exercises}\label{app:sim}

We evaluate the proposed model and posterior simulation algorithms using two simulation exercises. Specifically, we simulate count time series of length $N=400$ from a zero-inflated Poisson random walk with zero-inflation probability $(1-\pi) = 0.1$. For each simulated dataset, we estimate three models: (i) a homoskedastic Poisson random walk without zero inflation (RW), (ii) a homoskedastic zero-inflated Poisson random walk (ZI-RW), and (iii) a stochastic volatility specification with a zero-inflation mechanism (ZI-SV). For each model, we collect 75{,}000 posterior draws after a burn-in of 7{,}500 iterations.

\begin{figure}[p]
\centering

% -------------------- Panel A: Simulation 1 --------------------
\begin{subfigure}{\linewidth}
\centering
\begin{subfigure}[b]{0.32\linewidth}
\centering
\includegraphics[width=\linewidth]{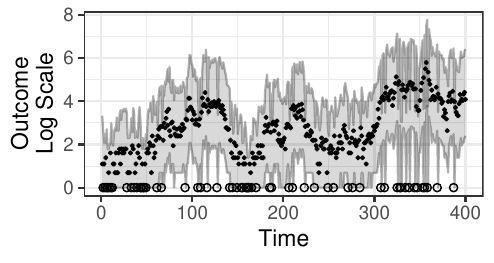}
\end{subfigure}
\begin{subfigure}[b]{0.32\linewidth}
\centering
\includegraphics[width=\linewidth]{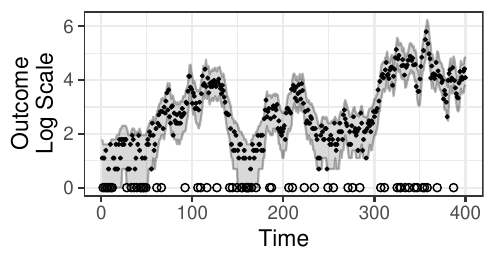}
\end{subfigure}
\begin{subfigure}[b]{0.32\linewidth}
\centering
\includegraphics[width=\linewidth]{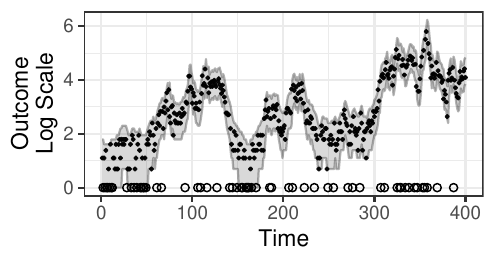}
\end{subfigure}

\begin{subfigure}[b]{0.32\linewidth}
\centering
\includegraphics[width=\linewidth]{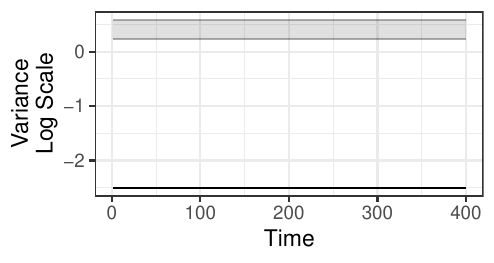}
\end{subfigure}
\begin{subfigure}[b]{0.32\linewidth}
\centering
\includegraphics[width=\linewidth]{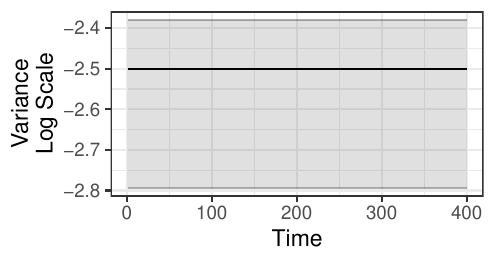}
\end{subfigure}
\begin{subfigure}[b]{0.32\linewidth}
\centering
\includegraphics[width=\linewidth]{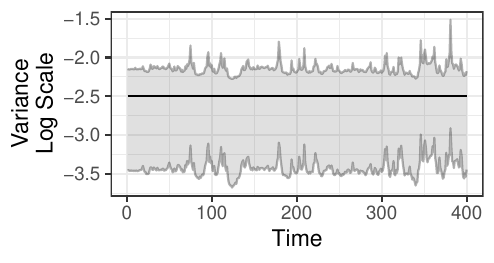}
\end{subfigure}
\caption{Simulation 1: Homoskedastic innovations.}
\label{fig:sim:hom}
\end{subfigure}

\vspace{1em}

% -------------------- Panel B: Simulation 2 --------------------
\begin{subfigure}{\linewidth}
\centering
\begin{subfigure}[b]{0.32\linewidth}
\centering
\includegraphics[width=\linewidth]{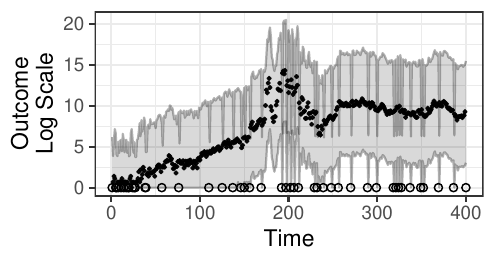}
\end{subfigure}
\begin{subfigure}[b]{0.32\linewidth}
\centering
\includegraphics[width=\linewidth]{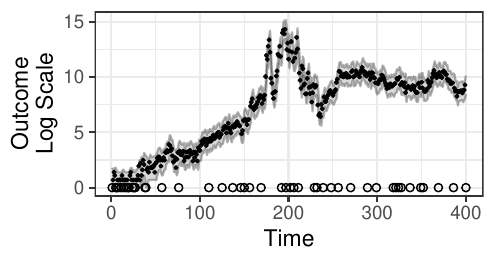}
\end{subfigure}
\begin{subfigure}[b]{0.32\linewidth}
\centering
\includegraphics[width=\linewidth]{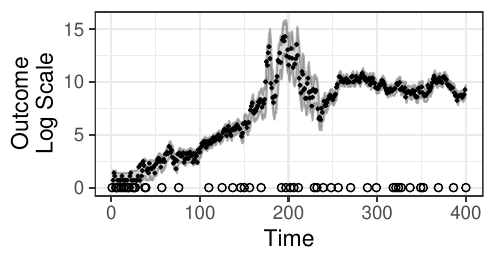}
\end{subfigure}

\begin{subfigure}[b]{0.32\linewidth}
\centering
\includegraphics[width=\linewidth]{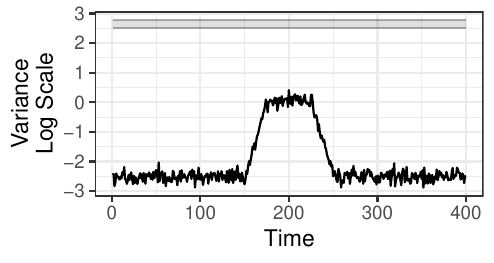}
\end{subfigure}
\begin{subfigure}[b]{0.32\linewidth}
\centering
\includegraphics[width=\linewidth]{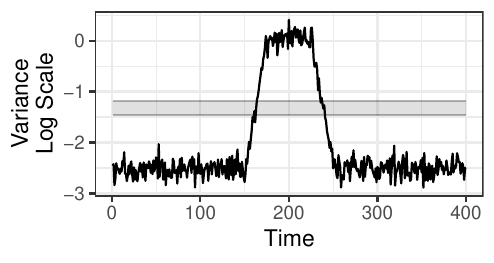}
\end{subfigure}
\begin{subfigure}[b]{0.32\linewidth}
\centering
\includegraphics[width=\linewidth]{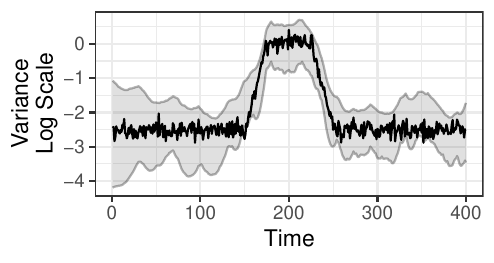}
\end{subfigure}
\caption{Simulation 2: Heteroskedastic innovations.}
\label{fig:sim:het}
\end{subfigure}

\caption{Simulation results. In each panel, the top row shows counts and fitted predictive distributions by time period on a logarithmic scale (grey shaded areas denote 90\% credible intervals; zero counts are highlighted separately on the zero line). The bottom row shows estimated 90\% credible intervals for $h_t$ (grey shaded) and true values (black line). Note that the $y$-axes differ across panels.}
\label{fig:sim}
\end{figure}

\paragraph*{Simulation 1: Homoskedastic Innovations.} In the first exercise, the true data-generating process features constant log-variance $h_t=-2.5$ for all $t$. Panel (a) of Fig.~\ref{fig:sim} summarizes the results. The correctly specified ZI-RW model recovers the parameters well. The ZI-SV model also captures the constant innovation variance accurately, albeit with somewhat wider credible intervals due to the additional flexibility in the volatility process. By contrast, the naive Poisson RW model that ignores zero inflation severely overestimates the innovation variance due to misattributing structural zeros to large negative innovations in the latent process. This is reflected both in the inflated uncertainty of the fitted counts (top left panel) and in the upward-biased variance estimates (bottom left panel). The root mean squared error between the true structural-zero indicators and the estimated structural-zero probabilities is very similar for the homoskedastic (0.1172) and stochastic volatility (0.1166) specifications, and both models recover structural zeros with high accuracy.

\paragraph*{Simulation 2: Heteroskedastic Innovations.} The second exercise features time-varying volatility. The log-variance $h_t$ starts at $-2.5$, gradually increases to $0$ during a high-volatility regime, and then returns to $-2.5$, with additional i.i.d.\ Gaussian noise of variance $0.025$ added to each $h_t$. Panel (b) of Fig.~\ref{fig:sim} reports the results. The ZI-SV model recovers the underlying volatility dynamics accurately, with 90\% credible intervals covering the true $h_t$ process throughout. By contrast, the ZI Poisson RW estimates an average innovation variance that overstates volatility during tranquil periods and understates it during volatile periods, resulting in poorly calibrated predictive intervals. The naive Poisson RW again performs worst, severely overestimating innovation variance due to ignoring zero inflation. As in Simulation~1, the root mean squared errors between the true structural-zero indicators and the estimated structural-zero probabilities are very similar for the homoskedastic (0.0756) and stochastic volatility (0.0723) specifications. While the stochastic volatility model performs slightly better, both models recover structural zeros with high accuracy.

\clearpage
\section{Additional Tables and Figures}

\textcolor{white}{..}
\vspace{2em}

\begin{figure}[!h]
     \centering
     \begin{subfigure}[b]{0.49\linewidth}
         \centering
         \includegraphics[width=\linewidth]{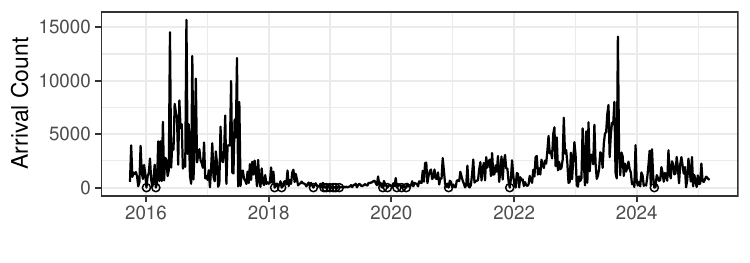}
     \end{subfigure}
          \begin{subfigure}[b]{0.49\linewidth}
         \centering
         \includegraphics[width=\linewidth]{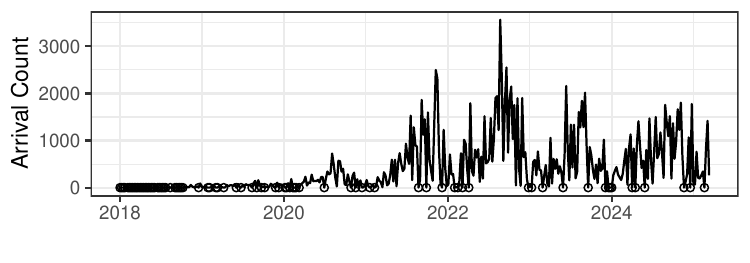}
     \end{subfigure}
     
        \caption{Sea border crossing case study datasets. Left: Mediterranean Sea border crossings to Italy by sample week from 2015W40 to 2025W11. Right: English Channel sea border crossings by sample week from 2018W01 to 2025W11. Panels display counts, with zero counts highlighted as dots on the zero line.}
        \label{fig:data_levels}
\end{figure}
\vspace{3em}

\begin{table}[!h]
  \centering
  \caption{Results of the Predictive Exercise: Full Sample.}
  \label{tab:fc_res_all}
  \adjustbox{width=\textwidth}{%
  \begin{threeparttable}
    \begin{tabular}{llrrrrrrrrrr}
      \toprule
      \multicolumn{2}{c}{} &
      \multicolumn{1}{c}{\textbf{LPS}} &
      \multicolumn{1}{c}{\textbf{RMSE}} &
      \multicolumn{1}{c}{\textbf{Corr.}} & 
\multicolumn{1}{c}{\textbf{$q_{01}$}} &
      \multicolumn{1}{c}{\textbf{$q_{05}$}} &
      \multicolumn{1}{c}{\textbf{$q_{10}$}} &
      \multicolumn{1}{c}{\textbf{$q_{90}$}} &
      \multicolumn{1}{c}{\textbf{$q_{95}$}} &
      \multicolumn{1}{c}{\textbf{$q_{99}$}} &
      \multicolumn{1}{c}{\textbf{n}}\\
      \midrule
ITA & Gaussian (excl. zeros) & -2180.957 & 7.558 & 0.370 & 0.032 & 0.068 & 0.092 & 0.900 & 0.932 & 0.980 & 250\\
ITA & Gaussian (incl. zeros) & -2203.042 & 7.559 & 0.372 & 0.016 & 0.028 & 0.052 & 0.932 & 0.956 & 0.984 & 250\\
ITA & Gaussian & -2166.997 & 7.483 & 0.345 & 0.000 & 0.016 & 0.076 & 0.900 & 0.932 & 0.980 & 250\\
ITA & Mixture & -2158.510 & 7.481 & 0.346 & 0.000 & 0.012 & 0.076 & 0.896 & 0.932 & 0.984 & 250\\
ITA & Student's t & -2160.850 & 7.481 & 0.346 & 0.000 & 0.012 & 0.072 & 0.892 & 0.928 & 0.984 & 250\\
ITA & Stoch. Vol & -2143.908 & 7.481 & 0.347 & 0.000 & 0.016 & 0.072 & 0.900 & 0.952 & 0.992 & 250\\
\addlinespace
UK & Gaussian (excl. zeros) & -2026.311 & 6.532 & 0.373 & 0.128 & 0.172 & 0.208 & 0.884 & 0.928 & 0.976 & 250\\
UK & Gaussian (incl. zeros) & -1956.947 & 6.577 & 0.337 & 0.028 & 0.088 & 0.104 & 0.896 & 0.916 & 0.960 & 250\\
UK & Gaussian & -1817.817 & 6.503 & 0.383 & 0.000 & 0.000 & 0.004 & 0.876 & 0.920 & 0.976 & 250\\
UK & Mixture & -1816.518 & 6.503 & 0.383 & 0.000 & 0.000 & 0.004 & 0.868 & 0.916 & 0.976 & 250\\
UK & Student's t & -1816.589 & 6.503 & 0.384 & 0.000 & 0.000 & 0.008 & 0.864 & 0.916 & 0.976 & 250\\
UK & Stoch. Vol & -1812.560 & 6.502 & 0.384 & 0.000 & 0.000 & 0.004 & 0.872 & 0.932 & 0.980 & 250\\
      \bottomrule
    \end{tabular}
    \begin{tablenotes}
      \footnotesize
      \item Note: Columns report the cumulative log predictive score (LPS), root mean square error (RMSE) on the logarithmic scale, correlation of forecasts and true values (Corr.), and empirical coverage with respect to the predictive 1st, 5th, 10th, 90th, 95th, and 99th quantiles ($q_{01}-q_{99}$), with $n$ denoting the number of hold-out observations. Results are summaries across out-of-sample periods. Second column refers to the innovation model for the Poisson random walk. `Gaussian (excl. zeros)' is a homoskedastic model where $s_t=0$ is fixed for all observations with $y_t=0$. `Gaussian (incl. zeros)' is a homoskedastic model where $s_t=1$ is fixed for all observations with $y_t=0$. All other models use an explicit zero inflation mechanism to estimate $s_t$.    \end{tablenotes}
  \end{threeparttable}
}
\end{table}

\begin{table}[!htbp]
\centering
\caption{Results of Exploratory Regression Models}
\label{tab:regression-results}
\begin{threeparttable}
\begin{tabular}{lcccccc}
\toprule
& \multicolumn{3}{c}{Mediterranean} & \multicolumn{3}{c}{English Channel} \\
\cmidrule(lr){2-4}\cmidrule(lr){5-7}
& Log Variance & Logit P($s_t=0$) & Log Arrivals
& Log Variance & Logit P($s_t=0$) & Log Arrivals \\
\midrule %% TinyTableHeader
Wave height & \num{0.011}   & \num{0.774}   & \num{-1.019}  & \num{0.049}   & \num{2.017}   & \num{-1.333}  \\
& (\num{0.030}) & (\num{0.108}) & (\num{0.071}) & (\num{0.026}) & (\num{0.206}) & (\num{0.111}) \\
2016        & \num{0.189}   & \num{0.124}   & \num{0.969}   &                &                &                \\
& (\num{0.159}) & (\num{0.573}) & (\num{0.379}) &                &                &                \\
2017        & \num{0.645}   & \num{-0.220}  & \num{0.506}   &                &                &                \\
& (\num{0.159}) & (\num{0.572}) & (\num{0.378}) &                &                &                \\
2018        & \num{-0.125}  & \num{0.780}   & \num{-1.268}  &                &                &                \\
& (\num{0.159}) & (\num{0.572}) & (\num{0.379}) &                &                &                \\
2019        & \num{-0.392}  & \num{0.798}   & \num{-1.893}  & \num{-0.552}  & \num{-0.704}  & \num{1.978}   \\
& (\num{0.160}) & (\num{0.574}) & (\num{0.380}) & (\num{0.075}) & (\num{0.591}) & (\num{0.318}) \\
2020        & \num{-0.082}  & \num{0.692}   & \num{-1.185}  & \num{-0.589}  & \num{-1.120}  & \num{3.351}   \\
& (\num{0.158}) & (\num{0.568}) & (\num{0.376}) & (\num{0.075}) & (\num{0.589}) & (\num{0.317}) \\
2021        & \num{0.026}   & \num{0.027}   & \num{-0.013}  & \num{-0.059}  & \num{-1.687}  & \num{4.226}   \\
& (\num{0.159}) & (\num{0.571}) & (\num{0.378}) & (\num{0.075}) & (\num{0.591}) & (\num{0.318}) \\
2022        & \num{-0.355}  & \num{-0.082}  & \num{0.400}   & \num{-0.101}  & \num{-1.731}  & \num{4.874}   \\
& (\num{0.159}) & (\num{0.570}) & (\num{0.377}) & (\num{0.075}) & (\num{0.591}) & (\num{0.318}) \\
2023        & \num{-0.131}  & \num{-0.134}  & \num{0.889}   & \num{-0.382}  & \num{-1.736}  & \num{4.693}   \\
& (\num{0.159}) & (\num{0.571}) & (\num{0.378}) & (\num{0.075}) & (\num{0.591}) & (\num{0.318}) \\
2024        & \num{0.328}   & \num{0.090}   & \num{-0.139}  & \num{-0.448}  & \num{-1.478}  & \num{4.751}   \\
& (\num{0.159}) & (\num{0.571}) & (\num{0.378}) & (\num{0.075}) & (\num{0.591}) & (\num{0.318}) \\
2025        & \num{-0.745}  & \num{-0.360}  & \num{0.477}   & \num{-0.253}  & \num{-2.171}  & \num{4.882}   \\
& (\num{0.214}) & (\num{0.771}) & (\num{0.510}) & (\num{0.131}) & (\num{1.035}) & (\num{0.557}) \\
Feb.        & \num{-0.205}  & \num{0.974}   & \num{-0.521}  & \num{-0.089}  & \num{-0.197}  & \num{0.273}   \\
& (\num{0.111}) & (\num{0.398}) & (\num{0.263}) & (\num{0.093}) & (\num{0.733}) & (\num{0.395}) \\
Mar.        & \num{-0.016}  & \num{0.502}   & \num{-0.338}  & \num{0.033}   & \num{-0.228}  & \num{0.256}   \\
& (\num{0.112}) & (\num{0.401}) & (\num{0.266}) & (\num{0.093}) & (\num{0.738}) & (\num{0.397}) \\
Apr.        & \num{-0.058}  & \num{0.126}   & \num{-0.064}  & \num{0.021}   & \num{1.098}   & \num{-0.552}  \\
& (\num{0.114}) & (\num{0.411}) & (\num{0.272}) & (\num{0.098}) & (\num{0.776}) & (\num{0.418}) \\
May         & \num{-0.535}  & \num{0.422}   & \num{0.081}   & \num{-0.117}  & \num{1.143}   & \num{-0.436}  \\
& (\num{0.117}) & (\num{0.422}) & (\num{0.279}) & (\num{0.101}) & (\num{0.797}) & (\num{0.429}) \\
Jun.        & \num{-0.783}  & \num{0.672}   & \num{-0.076}  & \num{-0.046}  & \num{1.718}   & \num{-0.459}  \\
& (\num{0.124}) & (\num{0.446}) & (\num{0.295}) & (\num{0.104}) & (\num{0.822}) & (\num{0.442}) \\
Jul.        & \num{-0.926}  & \num{0.800}   & \num{-0.156}  & \num{-0.052}  & \num{0.873}   & \num{-0.051}  \\
& (\num{0.123}) & (\num{0.444}) & (\num{0.294}) & (\num{0.102}) & (\num{0.809}) & (\num{0.435}) \\
Aug.        & \num{-1.027}  & \num{0.809}   & \num{-0.143}  & \num{-0.129}  & \num{0.928}   & \num{0.434}   \\
& (\num{0.124}) & (\num{0.446}) & (\num{0.295}) & (\num{0.101}) & (\num{0.801}) & (\num{0.431}) \\
Sep.        & \num{-0.732}  & \num{0.687}   & \num{0.073}   & \num{-0.077}  & \num{1.217}   & \num{0.368}   \\
& (\num{0.119}) & (\num{0.428}) & (\num{0.283}) & (\num{0.099}) & (\num{0.781}) & (\num{0.420}) \\
Oct.        & \num{-0.464}  & \num{0.194}   & \num{0.376}   & \num{-0.041}  & \num{-0.912}  & \num{1.125}   \\
& (\num{0.113}) & (\num{0.405}) & (\num{0.268}) & (\num{0.095}) & (\num{0.751}) & (\num{0.404}) \\
Nov.        & \num{-0.372}  & \num{0.240}   & \num{0.503}   & \num{0.166}   & \num{-1.430}  & \num{1.633}   \\
& (\num{0.112}) & (\num{0.401}) & (\num{0.265}) & (\num{0.095}) & (\num{0.753}) & (\num{0.405}) \\
Dec.        & \num{-0.295}  & \num{0.749}   & \num{-0.046}  & \num{0.132}   & \num{-0.043}  & \num{0.614}   \\
& (\num{0.110}) & (\num{0.396}) & (\num{0.262}) & (\num{0.094}) & (\num{0.742}) & (\num{0.399}) \\
Num.Obs.    & \num{494}     & \num{494}     & \num{494}     & \num{376}     & \num{376}     & \num{376}     \\
R2          & \num{0.460}   & \num{0.190}   & \num{0.596}   & \num{0.310}   & \num{0.290}   & \num{0.646}   \\
R2 Adj.     & \num{0.435}   & \num{0.152}   & \num{0.577}   & \num{0.273}   & \num{0.252}   & \num{0.627}   \\
\bottomrule
\end{tabular}

\begin{tablenotes}[flushleft]
\footnotesize
\item \textit{Notes:} OLS standard errors are reported in parentheses. Baseline years are 2015 (Mediterranean) and 2018 (English Channel). Baseline season is January. Arrivals include all zero observations and are transformed as $\log(1+x)$.
\end{tablenotes}
\end{threeparttable}
\end{table}

\begin{figure}
    \centering
   \includegraphics[width=\textwidth]{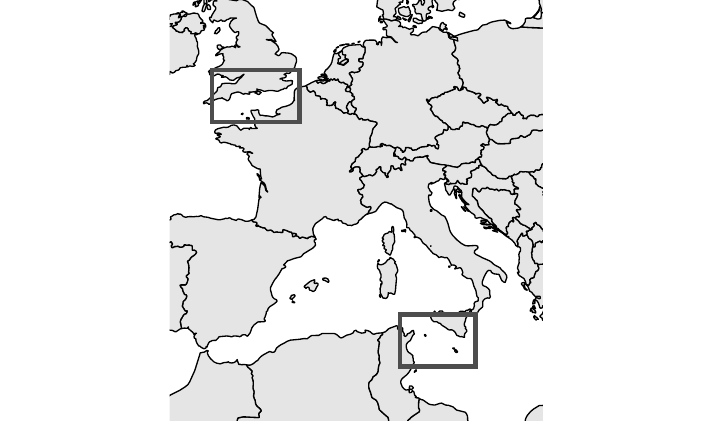}
   \caption{Map showing the boundary boxes for calculating weekly averages of wave height in the English Channel (upper box) and the Mediterranean Sea (lower box).}
    \label{fig:map}
\end{figure}

\begin{figure}[t]
     \centering
     \begin{subfigure}[b]{0.49\linewidth}
         \centering
         \includegraphics[width=\linewidth]{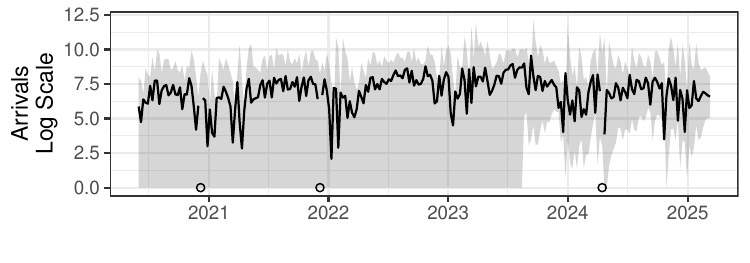}
     \end{subfigure}
          \begin{subfigure}[b]{0.49\linewidth}
         \centering
         \includegraphics[width=\linewidth]{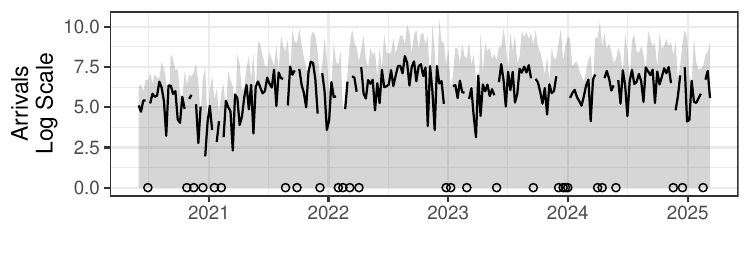}
     \end{subfigure}\\
     \begin{subfigure}[b]{0.49\linewidth}
         \centering
         \includegraphics[width=\linewidth]{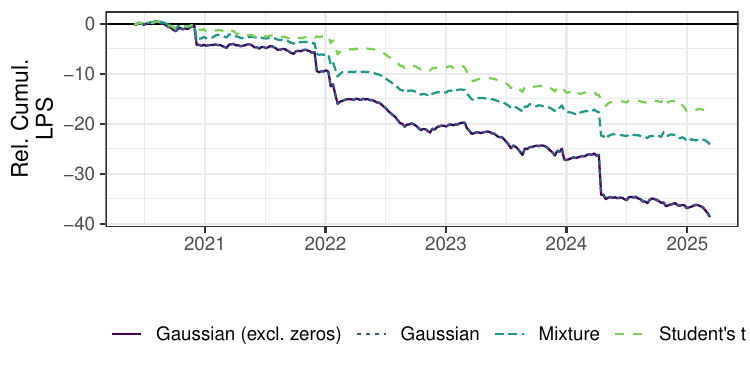}
     \end{subfigure}
     \begin{subfigure}[b]{0.49\linewidth}
         \centering
         \includegraphics[width=\linewidth]{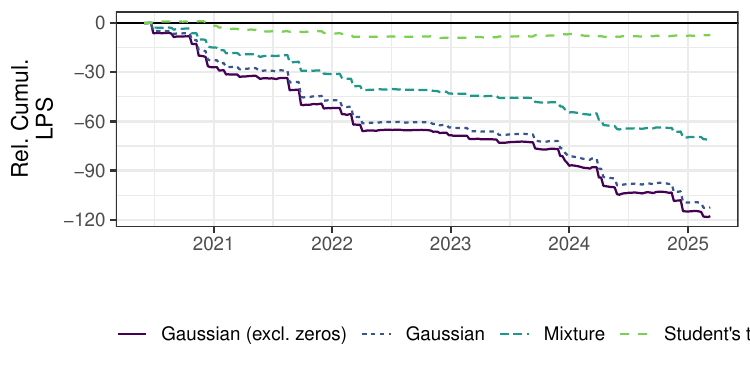}
     \end{subfigure}
     
        \caption{Top row: One-step-ahead predictive distributions and realized observations (black line) over the 250 hold-out periods. Zero observations are highlighted separately on the zero line.  Grey shaded areas are 90\% credible intervals.  Bottom row: Cumulative log predictive scores relative to the stochastic volatility model. Poisson model without zero inflation is not shown due to non-competitive performance. Left column: Mediterranean crossing data. Right column: English Channel crossing data. Hold-out period runs from 2020W23 to 2025W11. Results are marginal over $s_{T+1}$.}
        \label{fig:cumul-lps-full}
\end{figure}

\begin{figure}
    \centering
     \begin{subfigure}[b]{\linewidth}
         \centering
    \includegraphics[width=\linewidth]{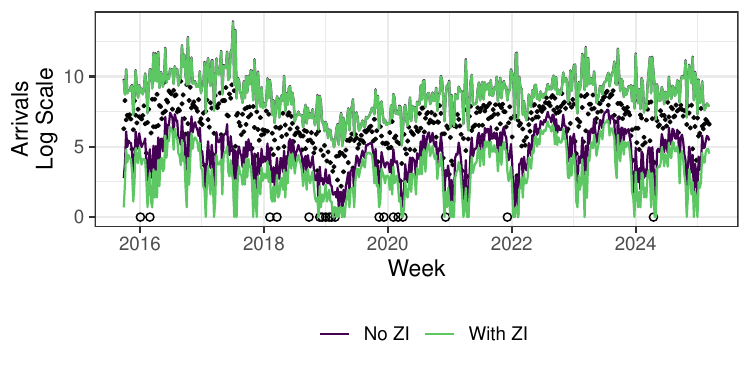}
     \end{subfigure}\\
     \begin{subfigure}[b]{\linewidth}
         \centering
         \includegraphics[width=\linewidth]{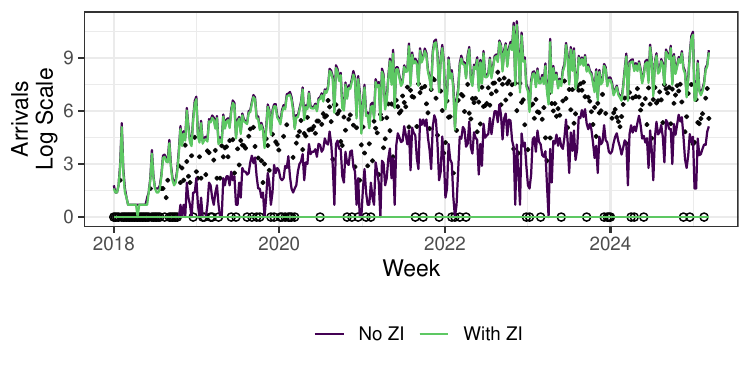}
     \end{subfigure}
\caption{Observed data (black dots) and estimated smoothed one-step-ahead predictive distributions over the sample period, summarized by the 5th and 95th percentiles (solid lines). The purple fit conditions on $s_t=1$ for all $t$ (i.e., predictions given no structural zero), whereas the green fit marginalizes over $s_t$. Top: Mediterranean Sea crossing data. Bottom: English Channel crossing data. ZI = zero-inflation.}
    \label{fig:both-fits}
\end{figure}

\end{document}